\RequirePackage{fix-cm}
\documentclass[twocolumn,epjc3]{svjour3}
\smartqed 
\RequirePackage{graphicx}
\usepackage{amsmath}
\usepackage{epstopdf}

\usepackage{xspace}

\usepackage[english]{babel}

\usepackage{amsfonts}
\usepackage{amssymb}
\usepackage{dsfont}
\usepackage[mathscr]{eucal}
\usepackage{braket}
\usepackage{cite}
\usepackage{titleref}
\usepackage{varioref}
\usepackage{enumitem}
\usepackage{mathtools}
\usepackage{tabularx}

\usepackage{epsfig}

\newcommand{\rmd}{\ensuremath{\mathrm{d}}}
\newcommand{\rme}{\ensuremath{\mathrm{e}}}
\newcommand{\rmi}{\ensuremath{\mathrm{i}}}

\DeclareMathOperator{\diag}{diag}

\newcolumntype{L}{>{\raggedright\arraybackslash}X}

\journalname{}
\begin{document}

\title{Evolution of confined quantum scalar fields in curved spacetime. Part I}

\subtitle{Spacetimes without boundaries or with static boundaries in a synchronous gauge}


\author{Luis C.\ Barbado\thanksref{e1,addr1,addr2}
\and
Ana L.\ B\'aez-Camargo\thanksref{e2,addr1}
\and
Ivette Fuentes\thanksref{e3,addr1,addr3}
}

\thankstext{e1}{e-mail: luis.cortes.barbado@univie.ac.at}
\thankstext{e2}{e-mail: lucia.baez@univie.ac.at}
\thankstext{e3}{e-mail: ivette.fuentes@nottingham.ac.uk}


\institute{Quantenoptik, Quantennanophysik und Quanteninformation, Fakult\"at f\"ur Physik, Universit\"at Wien, Boltzmanngasse 5, 1090 Wien, Austria \label{addr1}
\and
Institut f\"ur Quantenoptik und Quanteninformation, \"Osterreichische Akademie der Wissenschaften, Boltzmanngasse 3, 1090 Wien, Austria \label{addr2}
\and
School of Mathematical Sciences, University of Nottingham, University Park, Nottingham NG7 2RD, UK \label{addr3}
}

\date{}

\maketitle

\begin{abstract}
We develop a method for computing the Bogoliubov transformation experienced by a confined quantum scalar field in a globally hyperbolic spacetime, due to the changes in the geometry and/or the confining boundaries. The method constructs a basis of modes of the field associated to each Cauchy hypersurface, by means of an eigenvalue problem posed in the hypersurface. The Bogoliubov transformation between bases associated to different times can be computed through a differential equation, which coefficients have simple expressions in terms of the solutions to the eigenvalue problem. This transformation can be interpreted physically when it connects two regions of the spacetime where the metric is static. Conceptually, the method is a generalisation of Parker's early work on cosmological particle creation. It proves especially useful in the regime of small perturbations, where it allows one to easily make quantitative predictions on the amplitude of the resonances of the field, providing an important tool in the growing research area of confined quantum fields in table-top experiments. We give examples within the perturbative regime (gravitational waves) and the non-perturbative regime (cosmological particle creation). This is the first of two articles introducing the method, dedicated to spacetimes without boundaries or which boundaries remain static in some synchronous gauge.



\end{abstract}

\section{Introduction} \label{intro}

Quantum field theory in curved spacetime is the theory that studies the evolution of quantum fields which propagate in a classical general relativistic background geometry. To this date, the theory has been able to yield many quantitative theoretical predictions for physically relevant problems within its scope, such as cosmological particle creation~\cite{Parker1968, Parker1969}, the Unruh effect~\cite{Unruh1976} or Hawking radiation~\cite{Hawking1975}. The success in the study of those concrete well-known problems can be related to the use of different mathematical techniques and simplifications, adapted to the specific problem and which yield the computations tractable. For instance, in the study of phenomena such as Hawking radiation or the Unruh effect, the presence of horizons (or ``would-be horizons'' \cite{Barcelo2011a, Barbado2011}) allows to simplify the computations leading to a thermal spectrum for the quantum field, at least within most of the usual approaches to these phenomena. Also, the presence of symmetries like homogeneity or isotropy is convenient to obtain quantitative results on particle creation in cosmological models.

In the recent years, the interest to verify the theory experimentally has increased significantly\cite{Lin2008, Lin2015, Downes2011, Downes2013, Dragan2013, Bruschi2014testing, Wang2014quantum}. Thus far, an analogous of Hawking radiation in Bose-Einstein condensates has already been experimentally demonstrated~\cite{Steinhauer:2015saa} (although the result is not free of controversy, see for example \cite{leonhardt2018questioning}). In particular, and due to the great improvement of the precision of quantum measurements in table-top experiments, it is expected that the theoretical predictions on quantum fields confined in cavities, and under the effect of small changes in the background geometry or the non-inertial motion of the cavity, will be tested experimentally in the near future~\cite{Howl2018, Ahmadi2014, Tian2015relativistic, Wilson2011observation}. Consequently, new open problems within the framework of the theory are arising, related to the new concrete systems under study, and therefore new adapted mathematical techniques will be necessary to approach them.

In this work, we introduce a general method for computing the evolution of a confined quantum scalar field in a globally hyperbolic spacetime, by means of a time-dependent Bogoliubov transformation. As it will be clearly shown along the article, the method proves to be especially useful to address the new kind of problems that we just mentioned, related to quantum fields confined in cavities and undergoing small perturbations, since it provides very simple recipes for computing the resonance spectrum and sensibility of the field to a given perturbation of the background metric or the boundary conditions. However, we wish to emphasise that the method can be applied to any confined quantum scalar field, both real or complex, evolving as a result of any kind of regular changes in the background metric and/or the boundary conditions, just under some minor assumptions. This generality will also be illustrated through a concrete example of an application of the method outside the scope of small perturbations (in cosmological particle creation) that will be considered in the article.

The method is based on the foliation of the spacetime in spacelike Cauchy hypersurfaces using a time coordinate. Since the method is intended for confined fields, these hypersurfaces will be compact, although not necessarily with boundaries. The core idea is to construct a basis of modes naturally associated to each hypersurface, and then provide a differential equation in time for the Bogoliubov transformation between the modes associated to two hypersurfaces at different times. This way, the evolution of the field in time is not obtained by solving the Klein-Gordon equation, but rather by solving a differential equation for the time-dependent Bogoliubov transformation.

The core idea of transferring the dynamics from the mode functions to the Bogoliubov coefficients appears for the first time in the pioneer work by Parker~\cite{Parker1969}. In this sense, our work is a conceptual extension of Parker's work to the general case of any confined quantum scalar field (with minor assumptions). However, the key mathematical construction has been developed as a generalisation of the technique introduced in~\cite{Jorma2013}. That previous technique, although with the strong limitation of being applicable only in Minkowski spacetime (the influence in the evolution of the field coming from the motion of the cavity), has already proven to be of special practical use in the study of quantum fields confined in cavities. Benefiting from conformal invariance, extensions to a few other metrics were also considered~\cite{Sabin2014, Ahmadi2014, Bruschi2016towards, Lock2017, Lock2017relativistic}. Yet, the scope of that method was notoriously reduced, something which the present work overcomes.

We want to emphasise that our way of proceeding makes the method completely different to the Hamiltonian diagonalisation approach introduced for the problem of cosmological particle creation~\cite{Zel1972, Grib1976}. In such approach, the construction of bases of modes associated to each time is used to diagonalise the Hamiltonian of the quantum field by a time-dependent Bogoliubov transformation in the Fock representation, such that the Hamiltonian is diagonal in terms of particle operators at every moment; and then this diagonalised Hamiltonian is used to compute the evolution of a state in time directly in the Fock representation. Therefore, this approach introduces a different notion of particles \emph{at each instant of time.} Problems with this approach appear because the Fock representations for different times can be unitarily inequivalent~\cite{Fulling1979, Fulling1989, BD1984}; that is, because the particle interpretation in this framework is in general wrong. Conversely, the method introduced in this article does not intend to quantise the field with a Fock representation at all times during its evolution. The construction of the modes at each time will be just an intermediate mathematical tool used to compute a time-dependent Bogoliubov transformation. Such Bogoliubov transformation has the usual physical interpretation in terms of effects on quantum particles (mode-mixing and particle creation) only when it relates regions of spacetime where the Fock quantisation has a valid physical interpretation in terms of particles, in particular due to the presence of a timelike Killing field. The Bogoliubov transformation then unambiguously describes the evolution of the field between the two regions, since it allows to compute the scattering matrix between the bases of the two different Fock quantisations~\cite{BD1984,dewitt1975quantum,birrell1980analysis}. We provide strong physical arguments to explain why the Fock representations that we consider are unitary equivalent and physically meaningful.

Following the above discussion, we take the opportunity to stress that, for the quantisation of the field, we simply rely on canonical Fock quantisation in the regions where it is clearly physically meaningful. In other words, the method is of general \emph{practical} applicability to compute the scattering matrix between two physically valid Fock representations of the field, and it is indeed an extremely powerful tool for such purpose; while, on the other hand, we have no intention of providing a fully consistent field quantisation in any region of a general spacetime (see the approaches to such problem in e.g.~\cite{Haag:1963dh, Benini:2017dfw, Wald:2009uh, 2018arXiv180403434D, 2018arXiv180601572D}).

This article constitutes the first of two articles introducing the method. In this first part we consider spacetimes without boundaries, or with timelike boundaries that remain static in some synchronous gauge, as explained in the next section. Spacetimes with timelike boundaries which do not remain static in any synchronous gauge require an specific treatment, which due to its extension we leave for a second forthcoming article. The present article is organised as follows. In Sect.~\ref{stat} we state the general physical problem for which we will construct the method, introducing the background metric, the field theory and the different assumptions that we consider; and also define two important mathematical objects that we will use. The nuclear part of the article is Sect.~\ref{nuclear}. In this section we construct the basis of modes associated to each hypersurface of the foliation of the spacetime, and formally compute the time-dependent Bogoliubov transformation between the modes of two different hypersurfaces, for which we give a differential equation and a formal solution. We also discuss the physical meaning of both the modes and the transformations. In Sect.~\ref{res} we consider the particularly important case of small perturbations and resonances, obtaining especially simple recipes for its solution. As an example of application, we consider the problem of a confined quantum field in the presence of a gravitational wave. In Sect.~\ref{cosmo} we apply the method to an example of cosmological particle creation. Finally, in Sect.~\ref{conclu} we present the summary and conclusions. In addition, in \ref{conditions} we discuss in detail the different conditions that we require for the method to work. In \ref{F_quant} we describe how the method handles the problems in which modes with exponential time dependence or zero-frequency modes are present. In \ref{proof} we prove that the bases of solutions to the Klein-Gordon equation on which we build our method are complete and orthonormal. In \ref{bogo_comp} we provide the detailed computation of the differential equation provided in Sect.~\ref{nuclear}, and prove that the Bogoliubov transformations obtained satisfy the Bogoliubov identities. In \ref{invar} we further support and extend the results obtained on small perturbations. In \ref{recipes} we provide for convenience a summary of the useful formulae for the application of the method.

\section{Preliminaries}\label{stat}

\subsection{Statement of the problem}\label{statement}

We consider a globally hyperbolic spacetime~$(M, g)$ of dimension~$N+1$, possibly with timelike boundary~$\partial M$ \cite{Ake:2018dzz}. In this spacetime we define a Klein-Gordon scalar field~$\Phi$ satisfying the equation
\begin{equation}
g^{\mu \nu} \nabla_\mu \nabla_\nu \Phi - m^2 \Phi - \xi R \Phi = 0;
\label{klein-gordon}
\end{equation}
where~$m \geq 0$ is the rest mass of the field, $g^{\mu \nu}$ is the spacetime metric, $R$ its scalar curvature and~$\xi \in \mathds{R}$ is a coupling constant (we use natural units~$\hbar = c = 1$). In the case of the existence of boundaries, we impose one of the following four possible boundary conditions to the field:

\begin{enumerate}
	\item[a)] Dirichlet vanishing boundary conditions
\begin{equation}
\Phi(x) = 0, \quad x \in \partial M.
\label{dirichlet}
\end{equation}

	\item[b)] Neumann vanishing boundary conditions
\begin{equation}
n^\mu \nabla_\mu \Phi(x) = 0, \quad x \in \partial M;
\label{neumann}
\end{equation}
where~$n^\mu(x)$ is the normal vector to~$\partial M$.

	\item[c)] Robin vanishing boundary conditions
\begin{equation}
n^\mu \nabla_\mu \Phi(x) + \gamma \Phi (x) = 0, \quad x \in \partial M;
\label{robin}
\end{equation}
where~$\gamma \in \mathds{R} \setminus \{0\}$ is the boundary parameter. Later on, we will introduce the possibility that this parameter is replaced by a function.

	\item[d)] Mixed vanishing boundary conditions
	
	In this case, Dirichlet, Neumann and Robin vanishing boundary conditions apply each to complementary regions of the boundary. Being just a combination of the other cases, we mention them here as a possibility, but for simplicity we will not consider them explicitly when constructing the method.

\end{enumerate}

The types of boundary conditions considered are by far the most common ones in physical problems. We do not discard that the method could also apply to other less used conditions, but we simply do not approach such possibility within this work.

Thanks to the global hyperbolicity, it is always possible to construct a Cauchy temporal function~$t$ in the full spacetime~\cite{Ake:2018dzz}, which provides a foliation in Cauchy hypersurfaces~$\Sigma_t$ of constant time. We introduce the Klein-Gordon inner product between two solutions of (\ref{klein-gordon}), given by
\begin{multline}
\langle \Phi', \Phi \rangle := \\
- \rmi \int_{\Sigma_{\tilde{t}}} \rmd V_{\tilde{t}}\ \left[ \Phi'(\tilde{t}) \left. \partial_t \Phi(t)^* \right|_{t=\tilde{t}} - \Phi(\tilde{t})^* \left. \partial_t \Phi'(t) \right|_{t=\tilde{t}} \right];
\label{scalar_product}
\end{multline}
which, for convenience, we already evaluated at a given Cauchy hypersurface~$\Sigma_{\tilde{t}}$, with~$\rmd V_{\tilde{t}}$ being its volume element. Both in the absence of boundaries, or under the boundary conditions~(\ref{dirichlet}-\ref{robin}), this inner product is independent of~$\Sigma_{\tilde{t}}$.

Finally, we introduce the three conditions on the Cauchy hypersurfaces and the temporal function that we need to ensure the applicability of the method. These conditions are:

\begin{enumerate}
	\item[A.]
		The Cauchy hypersurfaces~$\Sigma_t$ must be compact.

	\item[B.]
		For any Cauchy hypersurface~$\Sigma_t$, the Cauchy problem for the Klein-Gordon equation~(\ref{klein-gordon}) must be well-posed; that is, given as initial conditions the value of the field and of its first derivative with respect to~$t$ at~$\Sigma_t$, there exists an unique solution to the Klein-Gordon equation in the whole spacetime satisfying these conditions.\footnote{Under the presence of boundaries, these initial conditions must be compatible with the boundary conditions at the intersection~$\partial \Sigma_t = \Sigma_t \cap \partial M$.}
		
	\item[C.]
		Using the temporal function as a coordinate, the metric should be written as
\begin{equation}
\rmd s^2 = - \rmd t^2 + h_{i j} (t, \vec{x}) \rmd x^i \rmd x^j,
\label{metric}
\end{equation}
where~$h_{i j} (t)$ is a regular Riemannian metric.\footnote{For simplicity in the notation, from here on we will omit the dependence of~$h_{i j} (t)$ on the spatial coordinates, but we are considering the most general case in which this metric can have any (regular) dependence on the coordinates. We will not write the explicit dependence on the spatial coordinates for other quantities either, except when necessary. Also, in~(\ref{metric}) for simplicity we are assuming that for each hypersurface~$\Sigma_t$ there is one coordinate chart~$(x^1,\ldots,x^N)$ that completely covers it. This might not be the case, but considering several coordinate charts would be straightforward and would not affect the construction of the method, as far as for all of them the metric can be written as in~(\ref{metric}).} This is called a synchronous gauge.
		
\end{enumerate}

The necessity of each condition will become clear when constructing the method. A detailed discussion on their physical meaning and the limitations they introduce can be found in \ref{conditions}. In this first article introducing the method, we are going to consider the problems for which~$M$ either has no boundary, or its boundary remains parallel to the gradient of the temporal function~$t$, that is, it remains static in some synchronous gauge. Problems for which there exists no adequate temporal function (satisfying the required conditions) such that the boundary remains parallel to its gradient will be considered in the second article.

\subsection{Space of functions over~$\Sigma_t$ and self-adjoint operator}\label{objects}

Let us introduce two mathematical objects that are pivotal for the method. First, we define~$\Gamma_t$ as a subspace of the space of square integrable smooth functions over a Cauchy hypersurface, $\Gamma_t \subseteq C^{\infty} (\Sigma_t) \cap L^2 (\Sigma_t)$. If~$M$ (and therefore~$\Sigma_t$) has no boundary, then $\Gamma_t = C^{\infty} (\Sigma_t) \cap L^2 (\Sigma_t)$. If there are boundaries, then~$\Gamma_t$ is the restriction of~$C^{\infty} (\Sigma_t) \cap L^2 (\Sigma_t)$ to functions satisfying the boundary conditions in~$\partial \Sigma_t$ compatible with the boundary conditions in~$\partial M$; which, since the boundaries are parallel to~$\partial_t$, read
\begin{align}
\Phi(\vec{x}) & = 0 \quad & \text{(Dirichlet)}, \label{dirichlet_static} \\
\vec{n} \cdot \nabla_{h(t)} \Phi(\vec{x}) & = 0 \quad & \text{(Neumann)}, \label{neumann_static} \\
\vec{n} \cdot \nabla_{h(t)} \Phi(\vec{x}) + \gamma \Phi(\vec{x}) & = 0 \quad & \text{(Robin)}; \label{robin_static}
\end{align}
for $\vec{x} \in \partial \Sigma_t$, where~$\vec{n}(\vec{x})$ is the normal vector to the boundary~$\partial \Sigma_t$ and~$\nabla_{h(t)}$ is the covariant derivative corresponding to the spatial metric~$h_{i j} (t)$. At this point, we mention that the boundary parameter~$\gamma$ could be also considered a function~$\gamma(\vec{x})$, but it must be independent of~$t$.\footnote{The justification for this apparently arbitrary limitation can be found in \ref{proof}.}

Second, we define the operator~$\hat{\mathscr{O}}(t)$ on~$\Gamma_t$ as
\begin{equation}
\hat{\mathscr{O}}(t) := - \nabla_{h(t)}^2 + \xi R^h(t) + m^2 + F(t);
\label{operator}
\end{equation}
where~$\nabla_{h(t)}^2$ is the Laplace-Beltrami differential operator, $R^{h}(t)$ the scalar curvature corresponding to the spatial metric~$h_{i j} (t)$ and $F(t) \geq 0$ a time-dependent non-negative quantity. The quantity~$F(t)$ is given by the condition that the operator~$\hat{\mathscr{O}}(t)$ at each time must be positive definite in the corresponding space~$\Gamma_t$ (any values of~$F(t)$ that meet this requirement would be valid). Notice that, thanks to condition~A and eventually to the boundary conditions, this operator is self-adjoint [with respect to the usual scalar product in $L^2 (\Sigma_t)$] and has a discrete spectrum with no accumulation points.\footnote{Assertions about the self-adjoint nature of the operator and about the properties of its spectrum should strictly be done over the extension of the operator defined on~$\Gamma_t$ to the full Hilbert space~$L^2 (\Sigma_t)$ (of which~$\Gamma_t$ is a dense subspace). This is a well-known mathematical procedure in the analysis of partial differential equations with elliptic operators and boundary value problems, which is the context in which we will make use of the operator. Therefore, we shall not get into the details of it. Another possibility is to directly consider the operator as acting on a Sobolev space over the Riemannian manifold. See for example~\cite{taylor1996partial}.} Moreover, if we fixed $F(t) = 0$ the operator would always have a minimum eigenvalue. Consequently, a finite value for~$F(t)$ such that the operator is positive definite can always be found. We refer to \ref{F_quant} for a discussion on the role of the quantity~$F(t)$, related to the presence of exponential time dependence of the solutions and of zero-frequency modes.\footnote{The discussion in \ref{F_quant} cannot however be fully understood until reading Sect.~\ref{nuclear}.} In many problems (such as the examples provided in this article) it is possible to fix $F(t) = 0$ while keeping~$\hat{\mathscr{O}}(t)$ positive definite, in which case the quantity can simply be ignored.

Once we have introduced the operator~$\hat{\mathscr{O}}(t)$, and making use of condition~C, we shall simplify the Klein-Gordon equation~(\ref{klein-gordon}) taking into account the form of the metric in~(\ref{metric}), obtaining
\begin{equation}
\partial_t^2 \Phi = - \hat{\mathscr{O}}(t) \Phi - q(t) \partial_t \Phi - \xi \bar{R}(t) \Phi + F(t) \Phi;
\label{klein-gordon_2}
\end{equation}
where
\begin{equation}
q(t) := \partial_t \log \sqrt{h(t)},
\label{change_factor}
\end{equation}
with~$h(t)$ being the determinant of the spatial metric~$h_{i j} (t)$ (it is a factor which depends on the change of the metric of the spacelike hypersurfaces with time), and $\bar{R}(t) := R(t) - R^{h}(t)$ is the part of the full scalar curvature of~$g_{\mu \nu}$ which depends on time derivatives, given by
\begin{equation}
\bar{R}(t) = 2 \partial_t q(t) + q(t)^2 - \frac{1}{4} [\partial_t h^{i j} (t)] [\partial_t h_{i j} (t)].
\label{bar_scalar}
\end{equation}
The key role of condition~C has been to yield equation~(\ref{klein-gordon_2}), in which all the spatial derivatives present are those in the Laplace-Beltrami operator contained in~$\hat{\mathscr{O}}(t)$.

\section{Construction of the method}\label{nuclear}

\subsection{Construction of the bases of modes}\label{bases}

For each Cauchy hypersurface~$\Sigma_{\tilde{t}}$ we will construct a set of modes~$\{\Phi^{[\tilde{t}]}_n (t)\}$ fulfilling the following two properties:

\begin{enumerate}
	\item[I.]
		They will form, together with their complex conjugates~$\{\Phi^{[\tilde{t}]}_n (t)^*\}$, a complete orthonormal basis of the space of global solutions to the Klein-Gordon equation~(\ref{klein-gordon}) with respect to the inner product~(\ref{scalar_product}). We stress that each mode of the basis is defined in the whole spacetime, the label~$[\tilde{t}]$ meaning only that we \emph{associate} it to the corresponding hypersurface. Specifically, it is in this hypersurface that we will set its initial conditions.
		
	\item[II.]
		In the regions where~$\partial_t$ behaves like a Killing field around the hypersurface~$\Sigma_{\tilde{t}}$ [that is, $h_{i j} (t)$ remains constant for a long enough period around~$t=\tilde{t}$], and where the function~$F(t)$ introduced in~(\ref{operator}) can be made to vanish, the modes will be those with positive frequency with respect to~$t$ in that region (the corresponding complex conjugates will be those with negative frequency).\footnote{Regions in which~$F(t)$ cannot be made to vanish, but can be made arbitrarily small, may also be considered. We refer to \ref{F_quant} for further details on this condition.} When this is not the case, the way we choose the modes can be thought just as an unambiguous recipe to associate an orthonormal basis of solutions to each hypersurface, even if no notion of positive frequency modes exists.
		
\end{enumerate}

\begin{figure}[h]
\begin{center}
\includegraphics[width=6cm]{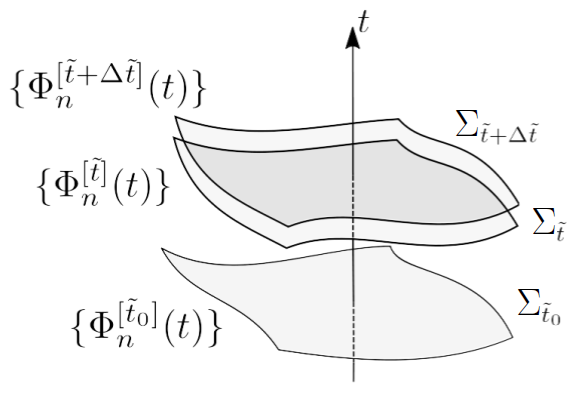}
\caption{Association of bases of modes~$\{\Phi^{[\tilde{t}]}_n (t)\}$ to Cauchy hypersurfaces~$\Sigma_{\tilde{t}}$.}
\label{slices}
\end{center}
\end{figure}

In Fig.~\ref{slices} we provide a graphical depiction of this association of bases of modes to Cauchy hypersurfaces, which shall be helpful when following the construction of the method. Since we impose that the modes are solutions to the Klein-Gordon equation, then because of condition~B the only quantities left to fully determine them are the initial conditions; which, as mentioned before, we are going to fix at~$\Sigma_{\tilde{t}}$. That is, we need to fix the quantities~$\Phi^{[\tilde{t}]}_n (\tilde{t})$ and~$\partial_t \Phi^{[\tilde{t}]}_n (t)|_{t=\tilde{t}}$ for each mode. The evaluations of the modes at~$t=\tilde{t}$, which will be functions in~$\Gamma_{\tilde{t}}$, are going to be given by the eigenvectors of the operator~$\hat{\mathscr{O}}(\tilde{t})$ in~(\ref{operator}):
\begin{equation}
\hat{\mathscr{O}}(\tilde{t}) \Phi^{[\tilde{t}]}_n (\tilde{t}) = (\omega^{[\tilde{t}]}_n)^2 \Phi^{[\tilde{t}]}_n (\tilde{t}).
\label{zeroth_derivative}
\end{equation}
Notice that $\tilde{t}$ in \eqref{zeroth_derivative} is just a parameter: This is a partial differential equation in the spatial coordinates of the given~$\Sigma_{\tilde{t}}$.\footnote{As we commented previously for the operator~$\hat{\mathscr{O}}(\tilde{t})$, this equation should also be posed in the full Hilbert space~$L^2 (\Sigma_t)$. It is known, however, that the functions representing the eigenvectors can be taken to belong to the dense subspace~$\Gamma_{\tilde{t}}$ \cite{taylor1996partial}.} Since~$\hat{\mathscr{O}}(\tilde{t})$ is positive definite, we can define the real and positive quantities~$\omega^{[\tilde{t}]}_n := +\sqrt{(\omega^{[\tilde{t}]}_n)^2}$ out of the eigenvalues of equation~(\ref{zeroth_derivative}). We also impose the following orthogonality and normalisation condition to the solutions:
\begin{equation}
\int_{\Sigma_{\tilde{t}}} \rmd V_{\tilde{t}}\ \Phi^{[\tilde{t}]}_n(\tilde{t})\Phi^{[\tilde{t}]}_m(\tilde{t})^* = \frac{\delta_{n m}}{2 \omega^{[\tilde{t}]}_n}.
\label{normalisation}
\end{equation}
Finally, the first time derivative of the modes evaluated at~$t=\tilde{t}$ will be given by
\begin{equation}
\left. \partial_t \Phi^{[\tilde{t}]}_n(t) \right|_{t=\tilde{t}} = -\rmi \omega^{[\tilde{t}]}_n \Phi^{[\tilde{t}]}_n(\tilde{t}).
\label{first_derivative}
\end{equation}

Equations~(\ref{zeroth_derivative}-\ref{first_derivative}) fully determine the initial conditions, and therefore their time evolution determines the modes~$\{\Phi^{[\tilde{t}]}_n, (\Phi^{[\tilde{t}]}_n)^*\}$. In~\ref{proof} we prove that this set of modes constitutes a complete orthonormal basis [in the Klein-Gordon inner product~(\ref{scalar_product})] of the space of solutions to the Klein-Gordon equation (satisfying the possible boundary conditions).

We have then proven that the modes fulfil property~I. Let us now consider property~II. One can easily realise that, when~$\partial_t$ behaves like a Killing field in some spacetime region~$S$ around~$\Sigma_{\tilde{t}}$, then $q(t) = \bar{R}(t) = 0$ in~$S$. Also, because of the time symmetry and the requisite that $F(t) = 0$ the eigenvalue problem~(\ref{zeroth_derivative}) will have the same solutions for all the hypersurfaces~$\Sigma_t$ within~$S$. Taking into account these facts, it is easy to check that the modes of the form
\begin{equation}
\Phi^{[\tilde{t}]}_n (t) = \Phi^{[\tilde{t}]}_n (\tilde{t}) \rme^{-\rmi \omega^{[\tilde{t}]}_n (t-\tilde{t})}
\label{modes_with_freq}
\end{equation}
satisfy both the initial conditions~(\ref{zeroth_derivative}-\ref{first_derivative}) in~$\Sigma_{\tilde{t}}$ and the Klein-Gordon equation~(\ref{klein-gordon_2}) in the whole region~$S$. Therefore, these modes correspond to the modes that we are actually assigning to the hypersurface~$\Sigma_{\tilde{t}}$. Now, if the interval of time that~$S$ embraces is large enough so as to explore the minimum frequency in the spectrum (now we can really call~$\omega^{[\tilde{t}]}_n$ a frequency), then it is clear that we can talk about the modes~(\ref{modes_with_freq}) as forming an orthonormal basis of modes with well-defined positive frequency with respect to~$t$. Property~II is therefore also fulfilled.

\subsection{Time-dependent Bogoliubov transformation}

Let us write down the Bogoliubov coefficients between any two orthonormal bases of modes, associated to the hypersurfaces~$\Sigma_{\tilde{t}_0}$ and~$\Sigma_{\tilde{t}}$:
\begin{align}
\alpha_{n m} (\tilde{t}, \tilde{t}_0) & := \langle \Phi^{[\tilde{t}]}_n, \Phi^{[\tilde{t}_0]}_m \rangle, \label{bogoliubov_a} \\
\beta_{n m} (\tilde{t}, \tilde{t}_0) & := - \langle \Phi^{[\tilde{t}]}_n, (\Phi^{[\tilde{t}_0]}_m)^* \rangle.
\label{bogoliubov_b}
\end{align}
Since, according to condition~B, all the modes are well defined globally, these coefficients exist and are unique as functions of time for all times. The objective of the method is to provide a differential equation in time that allows for their computation, without having to explicitly compute the time dependence of the modes.

For convenience, we write the Bogoliubov transformation in the matrix notation as
\begin{equation}
U(\tilde{t}, \tilde{t}_0) :=
\left(
\begin{array}{cc}
	\alpha (\tilde{t}, \tilde{t}_0) & \beta (\tilde{t}, \tilde{t}_0) \\
	\beta (\tilde{t}, \tilde{t}_0)^* & \alpha (\tilde{t}, \tilde{t}_0)^*
\end{array}
\right).
\label{bogoliubov_matrix}
\end{equation}
In \ref{main_calc} we prove that this time-dependent Bogoliubov transformation satisfies the differential equation
\begin{equation}
\frac{\rmd}{\rmd \tilde{t}} U(\tilde{t}, \tilde{t}_0) = \left[\rmi \Omega(\tilde{t}) + K(\tilde{t}) \right] U(\tilde{t}, \tilde{t}_0);
\label{bogoliubov_differential_equation}
\end{equation}
where we define the quantities
\begin{align}
\Omega (\tilde{t}) & := \diag (\omega^{[\tilde{t}]}_1, \omega^{[\tilde{t}]}_2, \ldots, -\omega^{[\tilde{t}]}_1, -\omega^{[\tilde{t}]}_2, \ldots), \nonumber \\
K(\tilde{t}) & :=
\left(
\begin{array}{cc}
	\hat{\alpha} (\tilde{t}) & \hat{\beta} (\tilde{t}) \\
	\hat{\beta} (\tilde{t})^* & \hat{\alpha} (\tilde{t})^*
\end{array}
\right);
\label{transform_matrices}
\end{align}
with
\begin{align}
\hat{\alpha}_{n m} (\tilde{t}) := &\ (\omega^{[\tilde{t}]}_m + \omega^{[\tilde{t}]}_n) \int_{\Sigma_{\tilde{t}}} \rmd V_{\tilde{t}}\ \left[ \frac{\rmd}{\rmd \tilde{t}} \Phi^{[\tilde{t}]}_n(\tilde{t}) \right] \Phi^{[\tilde{t}]}_m(\tilde{t})^* \nonumber \\
& + \int_{\Sigma_{\tilde{t}}} \rmd V_{\tilde{t}}\ \Phi^{[\tilde{t}]}_n(\tilde{t}) \left[\omega^{[\tilde{t}]}_n q(\tilde{t}) + \rmi \xi \bar{R}(\tilde{t}) \right] \Phi^{[\tilde{t}]}_m(\tilde{t})^* \nonumber \\
& + \frac{\delta_{n m}}{2 \omega^{[\tilde{t}]}_n} \left[ \frac{\rmd \omega^{[\tilde{t}]}_n}{\rmd \tilde{t}} - \rmi F(\tilde{t}) \right], \label{alpha_hat} \\
\hat{\beta}_{n m} (\tilde{t}) := &\ (\omega^{[\tilde{t}]}_m - \omega^{[\tilde{t}]}_n) \int_{\Sigma_{\tilde{t}}} \rmd V_{\tilde{t}}\ \left[ \frac{\rmd}{\rmd \tilde{t}} \Phi^{[\tilde{t}]}_n(\tilde{t}) \right] \Phi^{[\tilde{t}]}_m(\tilde{t}) \nonumber \\
& - \int_{\Sigma_{\tilde{t}}} \rmd V_{\tilde{t}}\ \Phi^{[\tilde{t}]}_n(\tilde{t}) \left[\omega^{[\tilde{t}]}_n q(\tilde{t}) + \rmi \xi \bar{R}(\tilde{t}) \right] \Phi^{[\tilde{t}]}_m(\tilde{t}) \nonumber \\
& - \left[ \frac{\rmd \omega^{[\tilde{t}]}_n}{\rmd \tilde{t}} - \rmi F(\tilde{t}) \right] \int_{\Sigma_{\tilde{t}}} \rmd V_{\tilde{t}}\ \Phi^{[\tilde{t}]}_n(\tilde{t}) \Phi^{[\tilde{t}]}_m(\tilde{t}). \label{beta_hat}
\end{align}

With the initial condition~$U(\tilde{t}_0, \tilde{t}_0) = I$, equation (\ref{bogoliubov_differential_equation}) has the formal solution
\begin{equation}
U(\tilde{t}_{\mathrm{f}}, \tilde{t}_0) = \mathscr{T} \exp \left\{ \int_{\tilde{t}_0}^{\tilde{t}_{\mathrm{f}}} \rmd \tilde{t} \left[ \rmi \Omega(\tilde{t}) + K(\tilde{t}) \right] \right\},
\label{bogoliubov_solution}
\end{equation}
where~$\mathscr{T}$ denotes time ordering.

Equations~(\ref{bogoliubov_differential_equation}) and~(\ref{bogoliubov_solution}) are the main result of this work: They switch from the time evolution of the modes to the time evolution of the transformation between bases. We also stress that one of the strengths of the method is that all the quantities appearing in~(\ref{transform_matrices}-\ref{beta_hat}), which are the coefficients of our differential equation, are known just by constructing the orthonormal basis of modes solutions to the eigenvalue equation~(\ref{zeroth_derivative}), for which the time~$\tilde{t}$ is just a parameter.

We can get rid of the trivial phase in~(\ref{bogoliubov_differential_equation}) and write the evolution in a more compact form. Apart from the phase introduced by the term~$\rmi \Omega(\tilde{t})$, one can check that the diagonal elements~$\hat{\alpha}_{n n} (\tilde{t})$ are purely imaginary [see relation~(\ref{bogos_condition}) in~\ref{id}]. Thus, we can get rid of the full trivial phase by defining the diagonal matrix
\begin{align}
\Theta(\tilde{t}) & := \exp \left\{ \int^{\tilde{t}} \rmd \tilde{t}' [\rmi \Omega (\tilde{t}') + A(\tilde{t}')] \right\}, \label{exponential_omega} \\
A(\tilde{t}) & := \diag(\hat{\alpha}_{1 1}, \hat{\alpha}_{2 2}, \ldots, -\hat{\alpha}_{1 1}, -\hat{\alpha}_{2 2}, \ldots); \nonumber
\end{align}
and writing the evolution in terms of a new operator~$Q(\tilde{t}, \tilde{t}_0)$ defined by
\begin{equation}
Q(\tilde{t}, \tilde{t}_0) := \Theta(\tilde{t})^* U(\tilde{t}, \tilde{t}_0).
\label{Q_definition}
\end{equation}
Replacing~(\ref{Q_definition}) in~(\ref{bogoliubov_differential_equation}), we get the differential equation [notice that~$\Theta(\tilde{t})^{-1} = \Theta(\tilde{t})^*$, since~$A(\tilde{t})^*=-A(\tilde{t})$]
\begin{align}
\frac{\rmd}{\rmd \tilde{t}} Q(\tilde{t}, \tilde{t}_0) = &\ \Theta(\tilde{t})^* \bar{K}(\tilde{t}) \Theta(\tilde{t}) Q(\tilde{t}, \tilde{t}_0), \label{Q_differential_equation} \\
\bar{K}(\tilde{t}) := &\ K(\tilde{t}) - A(\tilde{t});
\nonumber
\end{align}
which, again with the initial condition~$Q(\tilde{t}_0, \tilde{t}_0) = I$, has the formal solution
\begin{equation}
Q(\tilde{t}_{\mathrm{f}}, \tilde{t}_0) = \mathscr{T} \exp \left[ \int_{\tilde{t}_0}^{\tilde{t}_{\mathrm{f}}} \rmd \tilde{t}\ \Theta(\tilde{t})^* \bar{K}(\tilde{t}) \Theta(\tilde{t}) \right].
\label{Q_solution}
\end{equation}

In \ref{id} we check that the Bogoliubov transformations~(\ref{bogoliubov_solution}) and~(\ref{Q_solution}) obtained with this method satisfy the Bogoliubov identities.

\subsection{Physical interpretation}

The time-dependent Bogoliubov transformation that the method provides contains all the information necessary to compute the evolution of the field in time. However, as we advanced in the Introduction, we do not pretend to give a quantisation for each and every basis of modes that we have constructed at each time. It is only in those regions~$S$ with time symmetry, that we described at the end of the Subsect.~\ref{bases}, where we can proceed to the usual Fock quantisation of the field;\footnote{There is also the requisite that~$F(t) = 0$ to avoid negative eigenvalues (see \ref{F_quant}). For simplicity, we omit to mention this requisite explicitly along the rest of the Subsection.} that is, defining the corresponding Fock space with its vacuum state and creation and annihilation operators (ones the adjoints of the others in the case of a real field) associated to the mode decomposition given by the method, which in such region would be a decomposition in modes with well-defined frequency (for brevity, we do not expose this whole construction explicitly, see e.g.~\cite{BD1984}). We rely on the fact that, under the existence of a timelike Killing vector field, Fock representation gives the correct physical description of a field in terms of particles perceived by the family of observers following the integral trajectories of the timelike Killing vector field. The ``time-dependent Bogoliubov coefficients'' that we constructed in the previous Subsection can be interpreted as true Bogoliubov coefficients relating the annihilation and creation operators of different mode decompositions only when they connect two regions in time where this valid quantisation procedure can be done. When this is not the case, within the scope of this article they should be considered just as an intermediate computational tool. Whether there is still some physical interpretation of the intermediate times between two static regions will be studied in future works. The Bogoliubov coefficients at those intermediate times may well be related to field observables (such as the stress-energy tensor) instead of particle observables, since those remain physically meaningful in the absence of timelike Killing fields.

A concern that can still be raised is that, even if two Fock quantisations done in two different regions~$S_1$ and~$S_2$, in which~$\partial_t$ is a Killing field, are both valid and physically correct (which we know to be the case), they could eventually still be unitarily inequivalent. Unfortunately, we do not have a rigorous mathematical demonstration against such possibility occurring, but we can provide a solid physical argument for the cases considered with our method, taking into account that we only consider regular metrics. The condition for unitary inequivalence in the case of two quantisations related by a Bogoliubov transformation is that~$\sum_{n, m} |\beta_{n m}|^2 \to \infty$. This implies a divergent number of particles in one of the quantisations for any state of the field with a finite number of particles in the other quantisation. If the energies of the particles in each quantisation were not limited from below, a divergent number of particles could happen without a divergent amount of energy (the so-called \emph{infrared catastrophe}~\cite{strocchi}). But thanks to condition~A and the positive definiteness of~$\hat{\mathscr{O}}(t)$ a minimum non-zero energy always exists. Therefore, for the problems considered with this method unitarily inequivalent quantisations would imply that, for states with finite energy in one of the quantisations, the energy diverges according to the other quantisation. However, for physically valid quantisations this possibility is also discarded by the regularity of the metric and the compactness of the spatial hypersurfaces, since these two facts imply that only finite amounts of energy can be interchanged with the field in finite times. Therefore, for the problems considered with this method, physically valid quantisations must be also unitarily equivalent.

\section{Small perturbations and resonances}\label{res}

Let us consider now the case in which the spatial metric~$h_{ij} (t)$ only changes in time by a small perturbation around some constant metric~$h^0_{ij}$; that is,
\begin{equation}
h_{ij} (t) = h^0_{ij} + \varepsilon \Delta h_{ij} (t),
\label{perturbation_metric}
\end{equation}
where~$\varepsilon \ll 1$. Since, if boundaries exist, they remain parallel to~$\partial_t$, all the spacelike hypersurfaces will be identical, and for convenience we will call them~$\Sigma^0 := \Sigma_{\tilde{t}}$. We additionally require that~$F(t)$ remains~$O(\varepsilon)$, so that the solutions to the problem for~$\varepsilon = 0$ are modes with well-defined frequency. Later on, we will check that such value of~$F(t)$ actually does not contribute at all to the physically relevant result of resonances, and thus the function~$F(t)$ can be fully ignored in the small perturbations regime. Staying to first order in~$\varepsilon$, we can write the relevant quantities for computing the Bogoliubov coefficients as
\begin{align}
\Phi^{[\tilde{t}]}_n(\tilde{t}) & \approx \Phi^0_n + \varepsilon \Delta \Phi^{[\tilde{t}]}_n(\tilde{t}), & q(\tilde{t}) & \approx \varepsilon \Delta q(\tilde{t}), \nonumber \\
\omega^{[\tilde{t}]}_n & \approx \omega^0_n + \varepsilon \Delta \omega^{[\tilde{t}]}_n(\tilde{t}), & \bar{R}(\tilde{t}) & \approx \varepsilon \Delta \bar{R}(\tilde{t}), \label{perturbation_quantities} \\
& & F(\tilde{t}) & \approx \varepsilon \Delta F(\tilde{t});
\nonumber
\end{align}
where~$\Phi^0_n$ and~$\omega^0_n$ are the modes and frequencies solution to~(\ref{zeroth_derivative}) with~$h^0_{ij}$ as the metric. It is immediate that~$\hat{\alpha}_{nm} (\tilde{t})$ and~$\hat{\beta}_{nm} (\tilde{t})$ in~(\ref{alpha_hat}) and~(\ref{beta_hat}) are~$O(\varepsilon)$, since the quantities~$\rmd \Phi^{[\tilde{t}]}_n(\tilde{t})/\rmd \tilde{t}$, $\rmd \omega^{[\tilde{t}]}_n/\rmd \tilde{t}$, $q(\tilde{t})$, $\bar{R}(\tilde{t})$ and~$F(\tilde{t})$ are all~$O(\varepsilon)$. We can then write~$\hat{\alpha}_{n m} (\tilde{t}) \approx \varepsilon \Delta \hat{\alpha}_{n m} (\tilde{t})$ and~$\hat{\beta}_{n m} (\tilde{t}) \approx \varepsilon \Delta \hat{\beta}_{n m} (\tilde{t})$, with
\begin{align}
\Delta \hat{\alpha}_{n m} (\tilde{t}) := &\ (\omega^0_m + \omega^0_n) \int_{\Sigma^0} \rmd V^0\ \left[ \frac{\rmd}{\rmd \tilde{t}} \Delta \Phi^{[\tilde{t}]}_n(\tilde{t}) \right] (\Phi^0_m)^* \nonumber \\
& + \int_{\Sigma^0} \rmd V^0\ \Phi^0_n \left[\omega^0_n \Delta q(\tilde{t}) + \rmi \xi \Delta \bar{R}(\tilde{t}) \right] (\Phi^0_m)^* \nonumber \\
& + \frac{\delta_{n m}}{2 \omega^0_n} \left[ \frac{\rmd \Delta \omega^{[\tilde{t}]}_n}{\rmd \tilde{t}} - \rmi \Delta F (\tilde{t}) \right], \label{perturbation_alpha_hat} \\
\Delta \hat{\beta}_{n m} (\tilde{t}) := &\ (\omega^0_m - \omega^0_n) \int_{\Sigma^0} \rmd V^0\ \left[ \frac{\rmd}{\rmd \tilde{t}} \Delta \Phi^{[\tilde{t}]}_n(\tilde{t}) \right] \Phi^0_m \nonumber \\
& - \int_{\Sigma^0} \rmd V^0\ \Phi^0_n \left[\omega^0_n \Delta q(\tilde{t}) + \rmi \xi \Delta \bar{R}(\tilde{t}) \right] \Phi^0_m \nonumber \\
& - \left[ \frac{\rmd \Delta \omega^{[\tilde{t}]}_n}{\rmd \tilde{t}} - \rmi \Delta F (\tilde{t}) \right] \int_{\Sigma^0} \rmd V^0\ \Phi^0_n \Phi^0_m; \label{perturbation_beta_hat}
\end{align}
where~$\rmd V^0$ is the volume element of the metric~$h^0_{ij}$. Later on, we will further simplify these expressions for practical purposes.

Let us obtain the Bogoliubov coefficients to order~$\varepsilon$ in terms of these expressions. In the case of small perturbations, it is more convenient to consider the evolution as expressed by the Bogoliubov transformation $Q(\tilde{t}_{\mathrm{f}}, \tilde{t}_0)$ in~(\ref{Q_solution}). Noticing that~$\bar{K}(\tilde{t})$ is~$O(\varepsilon)$, we only need to consider the zeroth order in the matrix~$\Theta(\tilde{t})$. To first order in~$\varepsilon$, the transformation reads
\begin{equation}
Q(\tilde{t}_{\mathrm{f}}, \tilde{t}_0) \approx I + \varepsilon \int_{\tilde{t}_0}^{\tilde{t}_{\mathrm{f}}} \rmd \tilde{t}\ \Theta^0(\tilde{t})^* \Delta \bar{K}(\tilde{t}) \Theta^0(\tilde{t});
\label{perturbation_Q}
\end{equation}
where
\begin{align}
\Theta^0(\tilde{t}) := &\ \diag (\rme^{\rmi \omega^0_1 \tilde{t}}, \rme^{\rmi \omega^0_2 \tilde{t}}, \ldots, \rme^{-\rmi \omega^0_1 \tilde{t}}, \rme^{-\rmi \omega^0_2 \tilde{t}}, \ldots), \nonumber \\
\Delta \bar{K}(\tilde{t}) := &\
\left(
\begin{array}{cc}
	\Delta \hat{\alpha} (\tilde{t}) & \Delta \hat{\beta} (\tilde{t}) \\
	\Delta \hat{\beta} (\tilde{t})^* & \Delta \hat{\alpha} (\tilde{t})^*
\end{array}
\right) \nonumber \\
& - \diag (\Delta \hat{\alpha}_{1 1}, \Delta \hat{\alpha}_{2 2}, \ldots, -\Delta \hat{\alpha}_{1 1}, -\Delta \hat{\alpha}_{2 2}, \ldots).
\label{perturbation_matrices}
\end{align}

We can show the resonance behaviour of the field in a clear way if we write explicitly the expressions for the Bogoliubov coefficients:
\begin{align}
\alpha_{n n} (\tilde{t}_{\mathrm{f}}, \tilde{t}_0) \approx &\ 1; \nonumber \\
\alpha_{n m} (\tilde{t}_{\mathrm{f}}, \tilde{t}_0) \approx &\ \varepsilon \int_{\tilde{t}_0}^{\tilde{t}_{\mathrm{f}}} \rmd \tilde{t}\ \rme^{-\rmi (\omega^0_n - \omega^0_m) \tilde{t}} \Delta \hat{\alpha}_{n m} (\tilde{t}), \label{perturbation_alpha} \\
& \ n \neq m; \nonumber \\
\beta_{n m} (\tilde{t}_{\mathrm{f}}, \tilde{t}_0) \approx &\ \varepsilon \int_{\tilde{t}_0}^{\tilde{t}_{\mathrm{f}}} \rmd \tilde{t}\ \rme^{-\rmi (\omega^0_n + \omega^0_m) \tilde{t}} \Delta \hat{\beta}_{n m} (\tilde{t}). \label{perturbation_beta}
\end{align}

One can see that, in general, the Bogoliubov transformation will differ from the identity just by terms of first order in~$\varepsilon$, except for the cases where there are resonances. That is, if the perturbation of the metric contains some characteristic frequency~$\omega_{\mathrm{p}}$, then the same frequency will usually be also present in the quantities~$\Delta \hat{\alpha}_{n m} (\tilde{t})$ and~$\Delta \hat{\beta}_{n m} (\tilde{t})$; and if this frequency coincides with some difference between the frequencies of two modes, $\omega_{\mathrm{p}} = \omega^0_n - \omega^0_m$ (it is in resonance), then the corresponding coefficient~$\alpha_{n m} (\tilde{t}_{\mathrm{f}}, \tilde{t}_0)$ will grow linearly with the time difference~$\tilde{t}_{\mathrm{f}} - \tilde{t}_0$, and after enough time it will overcome the~$O(\varepsilon)$. Respectively, if the characteristic frequency coincides with some sum between the frequencies of two modes, $\omega_{\mathrm{p}} = \omega^0_n + \omega^0_m$, then the corresponding coefficient~$\beta_{n m} (\tilde{t}_{\mathrm{f}}, \tilde{t}_0)$ will grow linearly in time and eventually overcome the~$O(\varepsilon)$. For completeness, in \ref{invar1} we show that the resonances remain stable under small deviations of the frequency of the perturbation from the exact resonant frequency.

If the Fourier transform~$\mathscr{F}$ of~$\Delta \hat{\alpha}_{n m} (\tilde{t})$ [respectively $\Delta \hat{\beta}_{n m} (\tilde{t})$] exists as a well-defined function, which necessarily implies that the perturbation vanishes fast enough in the asymptotic past and future, another way to consider the resonances is by taking the limits~$\tilde{t}_0 \to -\infty$ and~$\tilde{t}_{\mathrm{f}} \to \infty$ in~(\ref{perturbation_alpha}) and~(\ref{perturbation_beta}) and writing
\begin{align}
\alpha_{n n} (-\infty,\infty) \approx &\ 1; \nonumber \\
\alpha_{n m} (-\infty,\infty) \approx &\ \varepsilon \sqrt{2 \pi}\ \mathscr{F} [\Delta \hat{\alpha}_{n m}] (\omega^0_n - \omega^0_m), \label{fourier_alpha} \\
& \ n \neq m; \nonumber \\
\beta_{n m} (-\infty,\infty) \approx &\ \varepsilon \sqrt{2 \pi}\ \mathscr{F} [\Delta \hat{\beta}_{n m}] (\omega^0_n + \omega^0_m). \label{fourier_beta}
\end{align}
That is, the Bogoliubov coefficients between the asymptotic past and future are proportional to the Fourier transforms evaluated at the corresponding subtraction (respectively addition) of frequencies. Evidently, resonances will occur if the frequency spectrum is peaked around one or more of these values.

It is important to remark that, since there is no strict time symmetry [neither~$F(t)$ vanishes in general], strictly speaking there would be no notion of particles associated to modes with well-defined frequency, except if the perturbation really vanishes in certain regions. However, even while the perturbation is ongoing, the deviation from the time symmetry is of order~$\varepsilon$. Thus, in the resonance regime, when one or more coefficients grow beyond this order, this lack of symmetry can be neglected with a solid physical argument: The deviation from time symmetry is small, and in particular much smaller than the effects measured. Therefore, in the case of resonance one can safely say that the physical effects are taking place for particles associated to modes with well-defined frequency.

In \ref{invar2}, we further support the interpretation in the previous paragraph by proving that the results on resonances do not depend on the particular choice of the basis of modes to order~$\varepsilon$ that we have used to compute the evolution. In order to give such proof, in \ref{invar2} we also prove a result that is useful at this point to further simplify the expressions in~(\ref{perturbation_alpha_hat}) and~(\ref{perturbation_beta_hat}): Any contribution of the form
\begin{equation}
\frac{\rmd X(\tilde{t})}{\rmd \tilde{t}} - \rmi (\omega^0_n - \omega^0_m) X(\tilde{t}),
\label{no_res_terms_a}
\end{equation}
respectively
\begin{equation}
\frac{\rmd X(\tilde{t})}{\rmd \tilde{t}} - \rmi (\omega^0_n + \omega^0_m) X(\tilde{t});
\label{no_res_terms_b}
\end{equation}
where~$X(\tilde{t})$ can be any function of time, appearing in the expression of~$\Delta \hat{\alpha}_{n m} (\tilde{t})$, respectively~$\Delta \hat{\beta}_{n m} (\tilde{t})$, will not affect the resonance behaviour of the corresponding modes. In simple words, this happens because, if~$X(\tilde{t})$ contains the correct resonant frequency in its Fourier expansion, the contribution of the corresponding term of the expansion to~(\ref{no_res_terms_a}) [resp.~(\ref{no_res_terms_b})] vanishes. We refer to \ref{invar2} for the details. Taking into account this fact, the expressions in~(\ref{perturbation_alpha_hat}) and~(\ref{perturbation_beta_hat}) can be simplified to
\begin{align}
\Delta \hat{\alpha}_{n m} (\tilde{t}) \equiv &\ \rmi [(\omega^0_n)^2 - (\omega^0_m)^2] \int_{\Sigma^0} \rmd V^0\ \Delta \Phi^{[\tilde{t}]}_n(\tilde{t}) (\Phi^0_m)^* \nonumber \\
& + \int_{\Sigma^0} \rmd V^0\ \Phi^0_n \left[\omega^0_n \Delta q(\tilde{t}) + \rmi \xi \Delta \bar{R}(\tilde{t}) \right] (\Phi^0_m)^*, \label{alpha_hat_simpl} \\
\Delta \hat{\beta}_{n m} (\tilde{t}) \equiv &\ -\rmi [(\omega^0_n)^2 - (\omega^0_m)^2] \int_{\Sigma^0} \rmd V^0\ \Delta \Phi^{[\tilde{t}]}_n(\tilde{t}) \Phi^0_m \nonumber \\
& - \int_{\Sigma^0} \rmd V^0\ \Phi^0_n \left[\omega^0_n \Delta q(\tilde{t}) + \rmi \xi \Delta \bar{R}(\tilde{t}) \right] \Phi^0_m \nonumber \\
& - [ 2 \rmi \omega^0_n \Delta \omega^{[\tilde{t}]}_n - \rmi \Delta F(\tilde{t}) ] \int_{\Sigma^0} \rmd V^0\ \Phi^0_n \Phi^0_m; \label{beta_hat_simpl}
\end{align}
where the symbol~`$\equiv$' denotes the equivalence relation ``gives the same resonances as''. Since resonances are the only physically meaningful result to be obtained from this computation, given a concrete problem one can always use these simplified expressions to compute the sensibility of the field to each resonance, by identifying the corresponding term of the Fourier expansion or value of the Fourier transform.

It is useful to notice that, although in principle we have assumed the orthonormality of the modes $\Phi^{[\tilde{t}]}_n(t)$, for using the expressions in~(\ref{alpha_hat_simpl}) and~(\ref{beta_hat_simpl}) only the non-perturbed modes $\Phi^0_n$ need to be strictly orthogonal [with respect to the usual inner product in~$L^2(\Sigma^0)$] and with the correct normalisation given by~(\ref{normalisation}) to zeroth order. The modes $\Phi^{[\tilde{t}]}_n(\tilde{t})$ need only to satisfy the unavoidable orthogonality already guaranteed by being correct solutions (to order~$\varepsilon$) of the eigenvalue problem~(\ref{zeroth_derivative}); that is, they will necessarily be orthogonal (to order~$\varepsilon$) unless they have the same eigenvalue~$(\omega^{[\tilde{t}]}_n)^2$. It is easy to check that any change of order~$\varepsilon$ in the normalisation of the modes (or in the inner product between modes with the same eigenvalue) would not contribute to~(\ref{alpha_hat_simpl}) or~(\ref{beta_hat_simpl}), because the corresponding factor $(\omega^0_n)^2 - (\omega^0_m)^2$ would make it to vanish. Therefore, when solving a specific problem, one only needs to explicitly ensure orthogonality and properly normalise the modes to zeroth order.

Finally, we give other alternative simplified formulas for the computation of~$\Delta \hat{\alpha}_{n m} (\tilde{t})$ and~$\Delta \hat{\beta}_{n m} (\tilde{t})$, which justification is given in \ref{simplif}. These formulas have the great advantage that they only imply the modes~$\Phi^0_n$ and eigenvalues~$\omega^0_n$ of the static problem:
\begin{align}
\Delta \hat{\alpha}_{n m} (\tilde{t}) & \equiv \int_{\Sigma^0} \rmd V^0\ [\hat{\Delta}(\tilde{t}) \Phi^0_n] (\Phi^0_m)^*, \label{super_simpl_a} \\
\Delta \hat{\beta}_{n m} (\tilde{t}) & \equiv - \int_{\Sigma^0} \rmd V^0\ [\hat{\Delta}(\tilde{t}) \Phi^0_n] \Phi^0_m; \label{super_simpl_b}
\end{align}
where~$\hat{\Delta}(\tilde{t})$ is a linear operator defined by its action on the basis~$\{\Phi^0_n\}$ as
\begin{align}
\hat{\Delta}(\tilde{t}) \Phi^0_n := &\ \big[ \rmi \Delta \hat{\mathscr{O}} (\tilde{t}) + \omega^0_n \Delta q(\tilde{t}) \nonumber \\
& + \rmi \xi \Delta \bar{R}(\tilde{t}) - \rmi \Delta F(\tilde{t}) \big] \Phi^0_n, \label{superoperator}
\end{align}
with~$\varepsilon \Delta \hat{\mathscr{O}} (\tilde{t})$ being the first order term in~$\varepsilon$ of the operator~$\hat{\mathscr{O}} (\tilde{t})$. For problems in which it is easy to compute the eigenvalues and eigenfunctions to first order in~$\varepsilon$, it might be easier to apply~(\ref{alpha_hat_simpl}) and~(\ref{beta_hat_simpl}); while for other problems the expressions in (\ref{super_simpl_a}-\ref{superoperator}) are definitely more convenient. Finally, as we already advanced, we can check in these last expressions that the quantity~$\Delta F(\tilde{t})$ never contributes to the resonances, since in~(\ref{superoperator}) its direct contribution to~$\hat{\Delta}(\tilde{t})$ cancels with its contribution through~$\Delta \hat{\mathscr{O}} (\tilde{t})$ [recall the definition in~(\ref{operator})].

The resonance can be consistently described in the regime of duration of the perturbation~$\Delta \tilde{t}$ such that~$1 \ll \omega_{\mathrm{p}} \Delta \tilde{t} \ll 1/\varepsilon$; since one needs the period of time to be reasonably larger than the inverse of the frequency being described, but on the other hand, one should keep the second order term in~$\varepsilon$ that we dropped in~(\ref{perturbation_Q}) significantly smaller than the first order term that we kept, so that the first order expansion of the exponential in~(\ref{Q_solution}) remains valid. Notice that, because of these limitations, the net effect after integrating in time under resonance might be of order greater than~$\varepsilon$, but still should be small as compared to unity. However, even the effect of such small Bogoliubov transformation could in principle be measured using the adequate probe states as the initial states of the field and quantum metrology techniques (see~\cite{Ahmadi2014, Sabin2014, Safranek2016, Safranek2015, Robbins2018}). Thus, even in these situations, in which the total energy involved was too small to produce a pair of particles or to promote an existing particle to a higher energy state, this would not mean that there are no measurable effects at all. Moreover, in this work we restrict ourselves to an idealised system in which the evolution of the field is perfectly unitary. More realistic systems may also present decoherence any time quanta is created or promoted, something that can then lead to cumulative effects.

\subsection{A concrete example}\label{gw}

In order to show the method in action, we will consider an example of application within the perturbative regime. In particular, we consider a field trapped in a cavity which is perturbed by a gravitational wave. This is a very interesting problem in the growing research field of gravitational wave detection with Bose-Einstein condensates. In~\cite{Sabin2014} the authors solve a similar problem by using the restricted version of the method in~\cite{Jorma2013}. They introduce a simplified toy model in one spatial dimension, taking advantage then of the conformal invariance in order to ``translate'' the problem into an equivalent one in Minkowski spacetime. With the general method presented here, it is possible to go beyond that scheme and solve the problem in three dimensions. The three-dimensional problem was also considered in~\cite{Robbins2018} with a very different approach. We will reproduce and extend the result in that work with very simple and straightforward calculations.

As in~\cite{Sabin2014} and~\cite{Robbins2018}, we consider our field to be the phonon field of a trapped Bose-Einstein condensate, therefore a real scalar massless quantum field. Following the work in~\cite{Visser:2010xv}, one can see that, in the case that the condensate remains stationary, the phonon field experiences an effective metric (with minimal coupling) which is the gravitational wave metric with the speed of light appearing in the component~$g_{00}$ replaced by the speed of sound in the condensate~$c_{\mathrm{s}}$. This speed of sound is the one that we normalise for this problem, $c_{\mathrm{s}} = 1$. We work in the TT-gauge and consider a wave with amplitude~$\varepsilon$ and frequency~$\Omega$ propagating in the $z$-direction and with polarisation in the $xy$-directions. This yields the metric
\begin{equation}
\rmd s^2 = - \rmd t^2 + [1 + \varepsilon \sin (\Omega t)] \rmd x^2 + [1 - \varepsilon \sin (\Omega t)] \rmd y^2 + \rmd z^2,
\label{metric_gw}
\end{equation}
where we have simplified~$\sin [\Omega (t - z/c)] \to \sin (\Omega t)$, as we have that $c \gg \Omega L_z$ (being~$L_z$ the size of the condensate in the $z$-direction), because of the orders of magnitude between the speed of light and the speed of sound. For simplicity, we consider that the condensate is trapped in a rectangular prism of lengths~$L_x$, $L_y$ and~$L_z$ aligned with the directions of propagation and polarisation of the wave. The boundaries are free-falling, and we impose Dirichlet vanishing boundary conditions.

For the metric in~(\ref{metric_gw}) and for $m = \xi = F(t) = 0$ the eigenvalue equation~(\ref{zeroth_derivative}) reads\footnote{When constructing the integration method, we have been using the different notations~$t$ and~$\tilde{t}$ for time, in order to distinguish between the evolution of modes in time and the choice of a concrete basis of modes associated to a Cauchy hypersurface~$\Sigma_{\tilde{t}}$. This was indeed necessary for constructing the method. But for its application to a concrete problem we will not need to explicitly consider the evolution in time~$t$ of a basis of modes anymore. Thus, we can relax the notation and replace~$\tilde{t} \to t$.}
\begin{multline}
\bigg[ -\frac{1}{1 + \varepsilon \sin (\Omega t)}\ \partial_x^2 -\frac{1}{1 - \varepsilon \sin (\Omega t)}\ \partial_y^2 \\
-\partial_z^2 \bigg] \Phi^{[t]}_{n m \ell} (t) = (\omega^{[t]}_{n m \ell})^2 \Phi^{[t]}_{n m \ell} (t),
\label{eq_gw}
\end{multline}
where it is clear that we will need three quantum numbers. Since the boundaries are free-falling, the boundary conditions read $\Phi(x=0) = \Phi(x=L_x) = 0$, and equivalently for the other dimensions. The solutions to order~$\varepsilon$, with the normalisation~(\ref{normalisation}) to zeroth order, are
\begin{align}
\Phi^{[t]}_{n m \ell} (t) & \approx \frac{2}{\sqrt{L_x L_y L_z \omega^0_{n m \ell}}} \sin (k^x_n x)\ \sin (k^y_m y)\ \sin (k^z_\ell z) \nonumber \\
& = \Phi^0_{n m \ell}, \quad n,m,\ell \in \mathds{N}^*;
\label{sols_gw}
\end{align}
where~$k^x_n := \pi n / L_x$, and equivalently for the other dimensions; with the eigenvalues to order~$\varepsilon$ given by
\begin{align}
\omega^{[t]}_{n m \ell} & \approx \omega^0_{n m \ell} + \varepsilon\ \frac{(k^y_m)^2 - (k^x_n)^2}{2\omega^0_{n m \ell}} \sin (\Omega t), \label{freqs_gw} \\
\omega^0_{n m \ell} & = \sqrt{(k^x_n)^2 + (k^y_m)^2 + (k^z_\ell)^2}.
\nonumber
\end{align}
Notice that, for this particular problem, the free-falling boundaries make the eigenvalues to depend on time while the modes are time-independent to order~$\varepsilon$. We plug the solutions into~(\ref{alpha_hat_simpl}) and~(\ref{beta_hat_simpl}), together with $\Delta q(t) = 0$ computed from~(\ref{change_factor}), obtaining
\begin{align}
\Delta \hat{\alpha}_{n m \ell}^{n'm'\ell'} (t) \equiv &\ \Delta \hat{\beta}_{n m \ell}^{n'm'\ell'} (t) \equiv 0, \label{no_diag_gw} \\
& \ (n', m', \ell') \neq (n, m, \ell); \nonumber \\
\Delta \hat{\beta}_{n m \ell}^{n m \ell} (t) \equiv &\ \rmi\ \frac{(k^x_n)^2 - (k^y_m)^2}{2 \omega_{n m \ell}^0} \sin (\Omega t). \label{beta_diag_gw}
\end{align}

By looking at these expressions, and considering the resonance behaviour found in~(\ref{perturbation_beta}), it is straightforward to obtain the conclusions: When $\Omega = 2 \omega^0_{n m \ell}$, the corresponding diagonal $\beta$-coefficient will grow linearly in time due to resonance as
\begin{equation}
\beta_{n m \ell}^{n m \ell} (t_{\mathrm{f}}, t_0) \approx \varepsilon\ \frac{(k^x_n)^2 - (k^y_m)^2}{4 \omega^0_{n m \ell}} (t_{\mathrm{f}} - t_0),
\label{Beta_res}
\end{equation}
while the rest of the coefficients will remain trivial in any case.

In~\cite{Robbins2018}, the authors consider the more realistic case in which the gravitational wave lasts some approximate time~$\tau$ and vanishes asymptotically, by replacing~$\sin (\Omega t) \to \rme^{-t^2/\tau^2} \sin (\Omega t)$ in the metric~(\ref{metric_gw}). We can easily handle this perturbation with our method by using the expression with the Fourier transform~(\ref{fourier_beta}), obtaining (for the only non trivial coefficients)
\begin{multline}
\beta_{n m \ell}^{n m \ell} (-\infty, \infty) \approx \varepsilon \sqrt{\pi}\ \frac{(k^x_n)^2 - (k^y_m)^2}{4 \omega^0_{n m \ell}}\ \tau \\
\times \left[\rme^{-(\Omega - 2 \omega^0_{n m \ell})^2 \tau^2/4} - \rme^{-(\Omega + 2 \omega^0_{n m \ell})^2 \tau^2/4}\right].
\label{Beta_Fourier}
\end{multline}
Equation~(\ref{Beta_Fourier}) exactly reproduces, through a very simple calculation (just a trivial eigenvalue problem and a trivial Fourier transform), the relation found in~\cite{Robbins2018} (equation~2.21). When~$\Omega \simeq 2 \omega^0_{n m \ell}$, it reproduces the resonance behaviour for the range of durations $1/\Omega \ll \tau \ll 1/|\Omega - 2 \omega^0_{n m \ell}|$, as discussed in \ref{invar1}.

We can obtain some further conclusions on this simple model. For example, it is straightforward to check that imposing Neumann vanishing boundary conditions leads to the same results. One would need to temporarily introduce a small rest mass term~$\Delta m$ in the field equation, in order to dodge the presence of a zero-frequency mode, as described in \ref{F_quant}; but it is easy to check that taking the limit~$\Delta m \to 0$ after solving the problem is in this case trivially well defined. Also, we can discuss what happens if we include the \emph{cross-polarisation} in the metric. This polarisation would add a term proportional to~$\partial_x \partial_y$ in the operator in~(\ref{eq_gw}). By using the expressions (\ref{super_simpl_a}-\ref{superoperator}), it is easy to check that this does not contribute to any Bogoliubov coefficient. Therefore, for waves coming perpendicular to any of the faces, the field is sensitive only to the polarisation which is aligned with the other faces. Again by using those expressions, it would not be difficult to even consider a wave propagating in an arbitrary direction, by applying the corresponding rotation matrices with the Euler angles to the metric, but for brevity we will not approach this problem here. In future publications~\cite{BarbadoWIP}, the authors will apply the method introduced in this article to study in detail the resonance properties of this and other models of trapping cavities, in what will be a fruitful application of the method to the study of gravitational wave detection with Bose-Einstein condensates.

\section{Application to cosmological particle creation}\label{cosmo}

In this section, we give an example of the application of the method beyond the perturbative regime. In particular, we use it to analyse a massive, minimally coupled Klein-Gordon field in a one-dimensional FLRW flat spacetime. The line element of this spacetime is given by
\begin{equation}
\rmd s^2=-\rmd t^2+a(t)^2\rmd x^2.
\label{FRW}
\end{equation}

For the metric in~(\ref{FRW}), where $h_{xx}(t)=a(t)^2$, and for $\xi = F(t) = 0$, the eigenvalue equation~(\ref{zeroth_derivative}) takes the form\footnote{As for the previous example in Subsect.~\ref{gw}, we replace~$\tilde{t} \to t$ in the notation.}
\begin{equation}
\label{eigeneq}
\left[-\frac{1}{a(t)^2} \partial_x^2 + m^2 \right]\Phi^{[t]}_n (t) = (\omega^{[t]}_n)^2 \Phi^{[t]}_n (t).
\end{equation}

As it is customary~\cite{Parker1969, BD1984}, we consider the spatial dimension to be a 1-torus, for which purpose we introduce periodic boundary conditions in space, so that~$\Phi(x=0) = \Phi(x=L)$, with~$L$ being the length of the torus. The solutions to~(\ref{eigeneq}), with the normalisation in~(\ref{normalisation}), are given by

\begin{equation}
\label{nmodes}
\Phi_n^{[t]}(t)=\frac{1}{\sqrt{2\omega_n^{[t]}|a(t)|L}}\ \rme^{\rmi k_n x}, \quad n \in \mathbb{Z};
\end{equation}
where $\omega_n^{[t]} = \sqrt{\frac{k_n^2}{a(t)^2}+m^2}$ and~$k_n := \frac{2\pi n}{L}$. From~(\ref{change_factor}) we get~$q(t)= \dot{a} (t) / a(t)$. Computing~(\ref{alpha_hat}) and~(\ref{beta_hat}) with the field modes~(\ref{nmodes}) we obtain
\begin{align}
\label{inf_bogoscosmo}
\hat{\alpha}_{nm}(t) & = 0; \quad \hat{\beta}_{nm} (t) = 0, \quad m\neq-n; \nonumber\\
\quad \hat{\beta}_{n(-n)} (t) & = -\frac{\dot{a}(t) m^2}{2 a(t) (\omega_n^{[t]})^2}.
\end{align}
Observe that if the scale factor is constant, then we obtain $\hat{\beta}_{n(-n)}(t)=0$ and the transformation is trivial. The transformation is also trivial when the scalar field is massless, which agrees with the fact that particles are created in $(1+1)$-dimensions only if a non-zero mass breaks the conformal invariance~\cite{BD1984}.

With the result obtained in~(\ref{inf_bogoscosmo}), the differential equation~(\ref{Q_differential_equation}) written directly for the Bogoliubov coefficients reads
\begin{align}
\dot{\alpha}_{nm} (t, t_0) & = \rme^{-2 \rmi\int^t \rmd t' \omega_n^{[t']}} \hat{\beta}_{n(-n)} (t) \beta_{(-n) m} (t, t_0)^*, \nonumber \\
\dot{\beta}_{nm} (t, t_0) & = \rme^{-2 \rmi\int^t \rmd t' \omega_n^{[t']}} \hat{\beta}_{n(-n)} (t) \alpha_{(-n) m} (t, t_0)^*.
\label{diff_eq_cosmo}
\end{align}
By replacing $n \to -n$ in the second equation, we realise that the system of equations can be decoupled in independent systems of two differential equations for each pair $\{\alpha_{nm},\beta_{(-n)m}\}$. Moreover, since the equations are homogeneous and the initial conditions are $\alpha_{nm}(t_0, t_0) = \delta_{nm}$ and $\beta_{nm}(t_0, t_0) = 0$, only the systems for which $m=n$ will have a non-trivial evolution, while the rest of the coefficients will all stay equal to zero. For the non-trivial systems, we can write down the two first order differential equations more conveniently as the following second order differential equation for $\alpha_{nn}(t, t_0)$:
\begin{multline}
\label{diffeq}
\ddot{\alpha}_{nn}(t,t_0)+\left[2 \rmi\ \omega_n^{[t]} - \frac{\dot{\hat{\beta}}_{n(-n)}(t)}{\hat{\beta}_{n(-n)}(t)} \right] \dot{\alpha}_{nn}(t,t_0) \\
-\hat{\beta}_{n(-n)}(t)^2 \alpha_{nn}(t,t_0)=0,
\end{multline}
with the initial conditions given by $\alpha_{nn}(t_0, t_0) = 1$ and $\dot{\alpha}_{nn}(t_0, t_0) \propto \beta_{(-n)n}(t_0, t_0)^* = 0$. Once computed the~$\alpha_{nn}(t,t_0)$ coefficients, computing another integral would give the $\beta_{(-n)n}(t,t_0)$ coefficients. However, it is much easier to compute them (up to an unimportant phase) out of the Bogoliubov identity~$\sum_m (|\alpha_{nm}|^2 - |\beta_{nm}|^2) = 1$, which in this case simplifies to
\begin{equation}
|\beta_{(-n)n}(t,t_0)|^2 = |\alpha_{nn}(t,t_0)|^2 - 1.
\label{betas_from_id}
\end{equation}

The mathematical results up to here are completely general, there is no assumption made on the form of the scale factor. As it was already discussed, the Bogoliubov coefficients computed will only have a clear physical interpretation when they connect regions in which~$\partial_t$ is a Killing field. We already reproduce the known result~\cite{Parker1969, BD1984} that in the problem considered there is never mode-mixing (the only non-zero~$\alpha$'s are the diagonal ones), but only particle creation in pairs of fully entangled particles with the same frequency and equal but opposite momentum. Finally, we point out that we could easily take the limit of an unconfined quantum field ($L \to \infty$) by simply replacing the role of the quantum number~$n$ by the wave number now raised to a continuum quantity, $k_n \to k \in \mathds{R}$; and the discrete frequencies by the corresponding dispersion relation, $\omega_n^{[t]} \to \omega^{[t]}(k)$.

\subsection{A concrete example}

To illustrate the previous general result, let us use it to compute numerically the Bogoliubov coefficients for a well-known toy model of cosmological expansion; in particular, the toy model appearing in \cite{BD1984}, pp.\ 59--62, and which was first proposed in~\cite{Bernard1977}. In this model, the expansion is explicitly written in terms of the conformal time~$\eta$, defined by~$\rmd \eta / \rmd t = 1/a(t)$. In this time, the scale factor is given by
\begin{equation}
a(\eta)=\sqrt{A+B\tanh(\rho \eta)},
\label{scale_factor}
\end{equation}
where $A$, $B$ and~$\rho$ are constants. One can see that, in the asymptotic past and future ($\eta, t \rightarrow \pm\infty$), the spacetime approaches the Minkowskian flat spacetime. In~\cite{BD1984} the Bogoliubov coefficients between the asymptotic regions~$\alpha_{nn} (\infty,-\infty)$ and~$\beta_{(-n)n} (\infty,-\infty)$ are obtained analytically by first solving the Klein-Gordon equation in conformal time through separation of variables.

\begin{figure}[h]
\begin{center}
\includegraphics[width=\columnwidth]{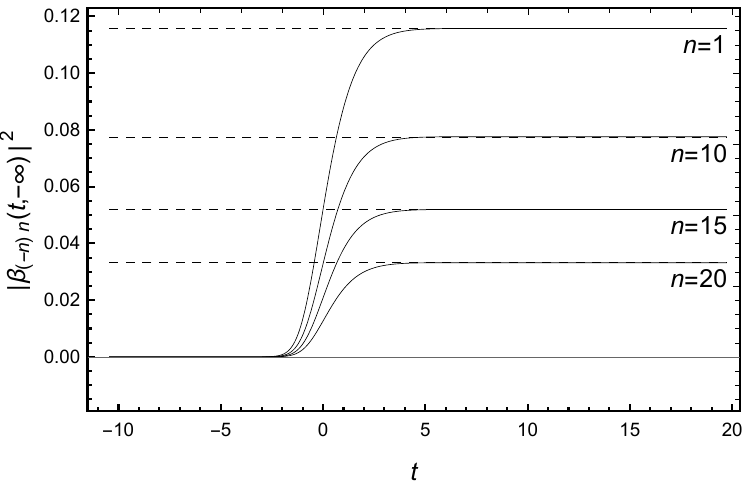}
\caption{Time evolution of~$|\beta_{(-n)n} (t,-\infty)|^2$ as obtained from numerical computation (solid lines), and asymptotic values~$|\beta_{(-n)n} (\infty,-\infty)|^2$ as given in~\cite{BD1984} (dashed lines). The mass of the field is~$m=0.1$, periodic boundary conditions are imposed for a length of~$L=1000$, and the parameters of~$a(\eta)$ are~$A=2.5$, $B=1.5$ and~$\rho = 1$.}
\label{graph}
\end{center}
\end{figure}

Unfortunately, for the given expression of~$a(\eta)$ we cannot obtain an explicit analytic expression for~$a(t)$ (one could expect this to happen, since the toy model is prepared \emph{ad hoc} to be solved in conformal time). Nonetheless, we can solve this problem numerically using the differential equation~(\ref{diffeq}) that we have obtained. We first compute the functions appearing in the coefficients of the equation, and finally solve the differential equation itself. The asymptotic initial conditions are~$\alpha_{nn} (-\infty,-\infty) = 1$ and~$\dot{\alpha}_{nn} (-\infty,-\infty) = 0$. Being trivially related by~(\ref{betas_from_id}), it is equivalent to follow the evolution of the~$\alpha$'s or the~$\beta$'s. For convenience, we consider the second ones, since they directly give the interesting result of particle creation for the quantum field. In Fig.~\ref{graph}, we plot a graphic with an example of numerical results for~$|\beta_{(-n)n} (t,-\infty)|^2$ for a concrete choice of the parameters in the problem, comparing them to the asymptotic values obtained analytically in~\cite{BD1984}.

One can clearly see that the asymptotic values in \cite{BD1984} are correctly reproduced. We also claim the attention about the fact that, even if the intermediate values of the coefficients cannot be assigned an immediate interpretation in terms of particle creation, the evolution of the coefficients in time is perfectly smooth and monotonically increasing, with a time dependence qualitatively similar to that of the scale factor itself. This ``well-behaved'' result is remarkable, and makes us to suspect that those intermediate values may have some physical meaning in terms of some sort of approximation, although the discussion of this idea lies beyond the scope of this article.

\section{Summary and conclusions}\label{conclu}

We have developed a method for computing the evolution of a confined quantum scalar field under general changes of the spacetime metric, under the assumptions given in Subsect.~\ref{statement}. In order to keep track of the evolution of the field, instead of solving the Klein-Gordon differential equation, the method computes a time-dependent Bogoliubov transformation between the bases of modes associated to different spacelike hypersurfaces. It provides the way to obtain the initial conditions of the basis of modes associated to each spacelike hypersurface as an eigenvalue problem in that hypersurface [equation~(\ref{zeroth_derivative})]. Once obtained these initial conditions, the method provides a system of homogeneous first order differential equations in time [equation~(\ref{bogoliubov_differential_equation}) or~(\ref{Q_differential_equation})] for the Bogoliubov transformation between the bases of modes associated to different times. The coefficients of such system of equations are computed out of the initial conditions of the modes themselves [equations~(\ref{transform_matrices}-\ref{beta_hat})]. Finally, we can also arrive to the formal generic solution of the system of equations [equation~(\ref{bogoliubov_solution}) or~(\ref{Q_solution})]. In this article we have considered the geometries without boundaries and those for which the boundaries remain static in some synchronous gauge, leaving the case of ``moving boundaries'' for a second article.

The time-dependent Bogoliubov transformation obtained with this method cannot, in general, be given a direct physical interpretation in terms of mode-mixing and particle creation phenomena, simply because the modes constructed for each spacelike hypersurface do not correspond (in general) to modes with well-defined frequency with respect to some time, and therefore they do not yield a physically unambiguous quantisation in terms of particles. However, if there is a region of spacetime (long enough in time) in which the time translation along the chosen time coordinate is a symmetry,\footnote{The requisite that~$F(t) = 0$ to avoid negative eigenvalues must be also satisfied. See \ref{F_quant}.} then automatically the modes as constructed with the method correspond to modes with well-defined frequency in that region. In such cases, the quantisation in these modes has the adequate interpretation in terms of particles, and the Bogoliubov transformation between two such regions has also its usual physical meaning in terms of mode-mixing and particle creation. In the regions where this is not the case, the modes and Bogoliubov transformations constructed can be considered just as an ``integration tool'' for computing the physically meaningful cases, although their potential meaning as some sort of approximation will be analysed in future works.

We stress that the method is of general applicability, just under the considered assumptions. However, as much as we stress this generality, we should also emphasise that the method proves to be of special practical use in the case of small perturbations of the metric (or the boundary conditions, within the next forthcoming article on the method) around a static situation. For clarity, we number here the reasons of this utility:

\begin{itemize}
	\item
		The perturbations of the modes can be obtained out of the eigenvalue equation~(\ref{zeroth_derivative}) just with perturbation theory to first order. In the same way, for computing the quantities~$\Delta \hat{\alpha} (\tilde{t})$ and~$\Delta \hat{\beta} (\tilde{t})$ one should only keep the first order in the perturbation, taking advantage of the equivalence relation with respect to resonances described in \ref{invar2} to further simplify the expressions [equations~(\ref{alpha_hat_simpl}) and~(\ref{beta_hat_simpl})]. Moreover, one can alternatively use the expressions (\ref{super_simpl_a}-\ref{superoperator}) to obtain the results directly from the solutions to the corresponding static problem.
		
	\item
		There is no need to explicitly solve the differential equation~(\ref{Q_differential_equation}) or to compute the exact transformation~(\ref{Q_solution}). In~(\ref{perturbation_alpha}) and~(\ref{perturbation_beta}) we can already see that, to first order, there is just a resonance behaviour. Thus, writing down the Fourier expansion of the perturbations~$\Delta \hat{\alpha} (\tilde{t})$ and~$\Delta \hat{\beta} (\tilde{t})$ automatically gives which Bogoliubov coefficients will stay to order~$\varepsilon$ (the order of the perturbation) and which (if any) will grow linearly in time due to resonance, and at which rate. Alternatively, when it is possible one can also take the Fourier transform to find the Bogoliubov coefficients between the asymptotic regions using~(\ref{fourier_alpha}) and~(\ref{fourier_beta}).
		
	\item
		Since the problem consist of a small perturbation of order~$\varepsilon$ around a static situation, in which there would be an exact notion of particles with well-defined frequency, the indefiniteness in such a notion will stay to order~$\varepsilon$. Thus, for the modes which are in resonance (the interesting ones), for which the Bogoliubov coefficients will grow beyond the order~$\varepsilon$, the physical interpretation in terms of mode-mixing and particle creation is perfectly valid.
	
	\item
		In~\ref{invar} we prove that the result on resonances is stable under small deviations from the resonant frequency. We also verify that the resonances obtained are the same regardless of which bases of modes to order~$\varepsilon$ one uses to integrate the evolution, which further supports that they are the physically meaningful result.
		
\end{itemize}

As an example of the application of the method, we have considered the perturbation of a confined field by gravitational waves in a realistic $3+1$~dimensional background. Through very simple calculations, we obtain the Bogoliubov coefficients, which agree with those given in~\cite{Robbins2018} for the same problem, and further discuss the role of the boundary conditions and the polarisation. The perturbative method could also prove its utility in other problems which are now under study, in which quantum systems are perturbed by small gravitational effects \cite{Howl2018, Ratzel2017testing, Ratzel2018dynamical}.

Finally, we have included an example of an application of the method to cosmological particle creation. We have obtained a differential equation that allows to compute the Bogoliubov coefficients for any one-dimensional FLRW flat spacetime, and applied this differential equation to an already known toy model of cosmological expansion, reproducing the correct results for the Bogoliubov coefficients between the asymptotic regions. In this way, we checked that the method can also be useful in problems beyond the perturbative regime. Out of the results obtained for the toy model, we highlight the smooth and monotonic behaviour in time of the Bogoliubov coefficients computed, even when they do not necessarily have an immediate physical interpretation. As we already mentioned, we leave for future work the analysis of the possible meaning of those values as some sort of approximation. We also leave for future work potential new applications of the integration method introduced in this article to other cosmological scenarios, as well as further applications in other problems in quantum field theory in curved spacetime.

\begin{acknowledgements}
The authors especially want to thank Jorma Louko for the rich exchange of ideas with us, which greatly helped solving many issues and significantly improved the article. We also want to thank Eugenia Colafranceschi, David E.\ Bruschi, Tupac Bravo, Daniel Hartley, Maximilian P.~E.\ Lock, Richard Howl, Joel Lindkvist, Jan Kohlrus, Dennis R\"atzel, Carlos Barcel\'o and Stephan Huimann for their useful comments and discussions during the elaboration of this article. We are equally grateful to Miguel S\'anchez Caja, Felix Finster and Simone Murro for clarifying our doubts on the mathematical background. Finally, we would like to thank an anonymous referee for some useful comments which helped to clarify the details of our mathematical construction. L.~C.~B.\ acknowledges the support from the research platform TURIS and from the European Commission via Testing the Large-Scale Limit of Quantum Mechanics (TEQ) (No.~766900) project, and from the Austrian-Serbian bilateral scientific cooperation no.\ 451-03-02141/2017-09/02. A.~L.~B.\ recognises support from CONACyT ref:579920/410674. I.~F.\ would like to acknowledge that this project was made possible through the support of the Penrose Institute, the grant ``Quantum Observers in a Relativistic World'' from FQXi's Physics of the Observer program, and the grant ``Leaps in cosmology: gravitational wave detection with quantum systems'' (No.~58745) from the John Templeton Foundation. The opinions expressed in this publication are those of the authors and do not necessarily reflect the views of the John Templeton Foundation.
\end{acknowledgements}

\appendix

\section{Discussion on the conditions for the method}\label{conditions}

In this Appendix we provide some further discussion on the conditions~A, B and~C introduced in Subsect.~\ref{statement}.

Condition~A reflects the fact that the method applies only for confined fields. It plays its role in ensuring that the operator~$\hat{\mathscr{O}}(t)$ in~(\ref{operator}) has a discrete spectrum, something crucial for the matrix formalism that we employ. In some situations (such as the example we give in Sect.~\ref{cosmo}), it might be possible to obtain results for unconfined fields by taking the limit of some typical length of the confinement region going to infinity after solving the problem~\cite{Parker1969, BD1984}. However, in most of the situations this limit may not be well defined, or may not commute with a limit previously taken when solving the problem, such as a late time limit.

Condition~B is introduced to ensure that the Cauchy problem for the scalar field is well posed, so that we can relate one-to-one initial conditions and solutions. Notice that this condition amounts to the predictability of the behaviour of the field in the given geometry. The aim of the method presented in this article is not to prove such predictability, but rather to compute the evolution of the field, for which it is reasonable to assume the predictability as given. If~$M$ has no boundaries, then condition~B is automatically ensured by the well-known theorems on existence and uniqueness of global solutions to partial differential equations. Unfortunately, there is to our knowledge no extension of these theorems in the presence of timelike boundaries. For Dirichlet boundary conditions, a proof of local well-posedness of the Cauchy problem for the Klein-Gordon equation is given in~\cite{pdes} (where the authors consider the uniform Lopatinski boundary condition, which includes Dirichlet as a specific case). Global well-posedness has been proven for the Dirac equation under MIT-boundary conditions, which are the analogous to Neumann boundary conditions for a scalar field~\cite{dirac}. In the case of Robin boundary conditions, global well-posedness is proven in~\cite{Dappiaggi:2018jsq} for the Klein-Gordon equation in static spacetimes of bounded geometry (for a wider class of boundary conditions including the Robin case). It is probable that global well-posedness of the Cauchy problem for the Klein-Gordon equation can be proven for the boundary conditions considered in this article as an extension of the results cited, but approaching such task is not in the aim of the present work, which goals are mainly computational. Therefore, lacking (so far) definitive theorems, we prefer to explicitly impose global well-posedness as a requisite. Nonetheless, we strongly believe that it will be satisfied, in the worst case, except for too stilted geometries, most probably without any physical interest.

Condition~C, which may seem unmotivated, is crucial for being able to write the Klein-Gordon equation as in~(\ref{klein-gordon_2}), with all the spatial derivatives collected in the self-adjoint operator~$\hat{\mathscr{O}}(t)$. Without this, the construction of the bases of modes and the time-dependent Bogoliubov transformation would not be possible, at least in the form we provide it. The existence of a gauge in which the crossed time-space terms in the metric cancel is guaranteed by the global hyperbolicity, with or without boundaries~\cite{Ake:2018dzz}. In most of the cases of physical interest, it should be also possible to make the lapse function~$-g_{tt}$ equal to one globally. In any case, this is always possible locally, so even if no global synchronous gauge exists, one may still apply the method ``by segments''. However, we cannot provide a general recipe on how to approach the joint of the results of the different segments.

\section{Function~$F(t)$, exponential time dependence and zero-frequency modes}\label{F_quant}

The quantity~$F(t)$ has been introduced in the operator~$\hat{\mathscr{O}}(t)$ in~(\ref{operator}) for a purely technical reason; namely, in order to have a positive-definite operator, so that all the eigenvalues obtained from the equation~(\ref{zeroth_derivative}) are positive, something necessary for the construction of bases of modes and the computation of Bogoliubov coefficients to hold as it has been exposed. Since, while there is no timelike Killing field, the construction of bases and Bogoliubov transformations is purely instrumental, there is no need to further discuss the role of~$F(t)$ in such case. However, when stating the properties of the bases constructed in Subsect.~\ref{bases}, in property~II, together with~$\partial_t$ being a timelike Killing field in a region of spacetime, we require also that $F(t) = 0$ in such region so that the modes constructed are modes with well-defined frequency (and we prove that this is the case). We can better understand this requisite, and in general the role of~$F(t)$, if we analyse what happens when~$\partial_t$ is a timelike Killing field but~$F(t) > 0$. There are two possibilities:

\begin{itemize}
	\item
		$F(t)$ must take some finite positive value, so that the operator~$\hat{\mathscr{O}}(t)$ is positive definite as required. If we insisted in setting $F(t) = 0$, we would obtain an operator with some negative eigenvalues. This would imply the existence of imaginary values for the quantities~$\omega^{[\tilde{t}]}_n$ obtained from the eigenvalue equation~(\ref{zeroth_derivative}). It is not difficult to check that in such case there exist fundamental solutions of the Klein-Gordon equation that have an exponential dependence on~$t$, by considering equation~(\ref{modes_with_freq}) with imaginary~$\omega^{[\tilde{t}]}_n$.\footnote{This situation should not be confused with the necessary presence of superradiance. See the Appendix in~\cite{Fulling1989}, or~\cite{PhysRevD.14.1939}.} Under this circumstance, it is not possible to proceed to the Fock quantisation of the field, even under the existence of a timelike Killing field. However, the method can still keep track of the evolution of the field in that region, if not for quantising it. By introducing the positive function~$F(t)$, we build bases of solutions associated to each hypersurface, although none of these solutions correspond to the fundamental solutions with exponential behaviour. They are just bases of solutions as instrumental as in the cases in which there is no timelike Killing field. But, precisely because they are not fundamental solutions, the Bogoliubov transformations between these different basis, obtained from the dynamics of the field, are not trivial, since $F(t)$ is present in~(\ref{alpha_hat}) and~(\ref{beta_hat}). In other words, the exponential behaviour that is ``artificially dropped'' from the modes by the introduction of~$F(t)$ is picked back and reproduced by the Bogoliubov transformation between them.

	\item
		$F(t)$ could be made arbitrarily small, but not zero, in order for the operator~$\hat{\mathscr{O}}(t)$ to be positive definite. In this case, setting $F(t) = 0$ would yield a positive semi-definite operator, which implies the existence of a fundamental zero-frequency mode. If one is not interested in the quantisation of the field in the region, one can proceed by giving~$F(t)$ some fixed value and compute the corresponding Bogoliubov coefficients. However, if one wishes to attempt quantisation in such region, one could instead add a small increment to the rest mass~$\Delta m$, so that the operator~$\hat{\mathscr{O}}(t)$ is positive definite even with $F(t) = 0$. Yet, after solving the problem one should take the limit~$\Delta m \to 0$ in order to recover the original field. This limit might not be trivial, and the method cannot give an \emph{a priori} prescription on how to take it. Even so, in Subsect.~\ref{gw} we give an example where this procedure proves to be successful in a trivial way. Notice however, that this difficulty with a zero-frequency mode is intrinsic to the field properties and not peculiar to the method.
		
\end{itemize}

We can identify some circumstances in which one can always set $F(t) = 0$ and forget about the quantity. In particular, let us consider the important case of a minimally coupled field ($\xi = 0$). In such case, non-zero rest mass and/or Dirichlet boundary conditions or Robin boundary conditions with~$\gamma > 0$~\cite{robinEigen} automatically make the operator~$\hat{\mathscr{O}}(t)$ to be positive definite; while for massless fields without boundaries or with Neumann boundary conditions the operator is positive semidefinite and could be handled with the procedure described in the second point above.

\section{Completeness and orthonormality of the bases of modes}\label{proof}

Let us prove that the set $\{\Phi^{[\tilde{t}]}_n, (\Phi^{[\tilde{t}]}_n)^*\}$ constructed in Subsect.~\ref{bases} is a complete basis of the space of solutions to the Klein-Gordon equation which satisfy the possible boundary conditions. Because of condition~B and the linearity of the Klein-Gordon equation, it is equivalent to prove that the set formed by the initial conditions for each mode in that set at~$\Sigma_{\tilde{t}}$ is itself a basis of the space of initial conditions at~$\Sigma_{\tilde{t}}$. If we recall~(\ref{first_derivative}), this set of initial conditions of the modes is given by the following set of pairs of values of the field and of its first time derivative at~$\Sigma_{\tilde{t}}$, $(\Phi(\tilde{t}), \partial_t \Phi(t) |_{t = \tilde{t}})$:
\begin{equation}
\{(\Phi^{[\tilde{t}]}_n (\tilde{t}), -\rmi \omega^{[\tilde{t}]}_n \Phi^{[\tilde{t}]}_n (\tilde{t})), (\Phi^{[\tilde{t}]}_n (\tilde{t})^*, \rmi \omega^{[\tilde{t}]}_n \Phi^{[\tilde{t}]}_n (\tilde{t})^*)\}.
\label{initial_cond_set}
\end{equation}

It is clear that the space of initial conditions at~$\Sigma_{\tilde{t}}$ [pairs of functions $(\Phi(\tilde{t}), \partial_t \Phi(t) |_{t = \tilde{t}})$] is~$L^2(\Sigma_{\tilde{t}}) \oplus L^2(\Sigma_{\tilde{t}})$. We will prove in short that the set in~(\ref{initial_cond_set}) is indeed a basis of such space. However, we also need to ensure that the elements in the basis satisfy the possible boundary conditions at~$\partial \Sigma_{\tilde{t}}$ compatible with the boundary conditions in~$\partial M$, when they exist. For the first component of the elements in~(\ref{initial_cond_set}) (the evaluation of the mode at~$\Sigma_{\tilde{t}}$), this compatibility means being in~$\Gamma_{\tilde{t}}$, since this is the definition of such subspace [together with the smoothness condition that simply picks up a representative function of the vector in the Hilbert space~$L^2(\Sigma_{\tilde{t}})$]. Such condition is clearly satisfied. Let us consider the second component of the elements in~(\ref{initial_cond_set}) (the time derivative of the mode at~$\Sigma_{\tilde{t}}$). If we take the total time derivative of any of the boundary conditions in~(\ref{dirichlet}-\ref{robin}) along the boundary, since the boundary itself remains static, we will easily find that the same boundary condition that applies to~$\Phi(t)$ must apply also to~$\partial_t \Phi(t)$. Therefore, we have that the second component of the elements in~(\ref{initial_cond_set}) must also belong to~$\Gamma_{\tilde{t}}$, something which is also clearly satisfied.

Observe that the fact that both the evaluation of the mode and of its first time derivative at some Cauchy hypersurface~$\Sigma_{\tilde{t}}$ satisfy the same boundary conditions is what allows for the construction of the modes as it has been done in this article [that is, imposing equation (\ref{first_derivative})]. This construction was therefore feasible thanks to the condition that the boundaries shall remain static in the synchronous gauge considered. If this is not the case and the boundaries are not parallel to the time flow~$\partial_t$, the boundary conditions at~$\partial M$ as given in~(\ref{dirichlet}-\ref{robin}) (or their derivatives along the boundary itself) mix spatial and time derivatives of the mode in the same equation, and the construction of initial conditions as presented in this article cannot be done any more. This is the situation that we will consider in the second forthcoming article on the method.\footnote{We can see at this point why, if we promote the boundary parameter~$\gamma$ of Robin boundary conditions to a function, a possibility given below equation~(\ref{robin_static}), such function must be time independent. Otherwise, the boundary conditions applying to the field and to its time derivative would also be different.}

Finally, let us prove that the set~(\ref{initial_cond_set}) is a basis of~$L^2(\Sigma_{\tilde{t}}) \oplus L^2(\Sigma_{\tilde{t}})$. For convenience, we temporarily simplify the notation for the functions and eigenvalues by replacing $\Phi^{[\tilde{t}]}_n (\tilde{t}) \to \Phi_n$ and $\omega^{[\tilde{t}]}_n \to \omega_n$. We notice first that clearly both sets of eigenvectors~$\{\Phi_n\}$ and~$\{\Phi_n^*\}$ are complete bases of~$L^2(\Sigma_{\tilde{t}})$.\footnote{Here~$L^2(\Sigma_{\tilde{t}})$ must be understood of course as a Hilbert space, that is, the space of square integrable functions modulo differences in a set of measure~$0$.} This means that the modes of the second basis can be expanded in the first basis with unique coefficients:
\begin{equation}
\Phi_n^* = \sum_m p_{n m} \Phi_m;
\label{expand_p}
\end{equation}
where
\begin{equation}
p_{n m} \neq 0 \Leftrightarrow \omega_n = \omega_m
\label{cond_p}
\end{equation}
since the subspaces generated by eigenfunctions of the same operator with different eigenvalues are orthogonal, and~$p_{n m}$ is clearly invertible. Also, any two arbitrary functions~$\Phi_A$ and~$\Phi_B$ in~$L^2(\Sigma_{\tilde{t}})$ can be expanded in the basis~$\{\Phi_n\}$ with unique coefficients:
\begin{equation}
\Phi_A = \sum_n a_n \Phi_n, \quad \Phi_B = \sum_n b_n \Phi_n.
\label{expand_a_b}
\end{equation}

Using the expansions in~(\ref{expand_p}) and~(\ref{expand_a_b}) we will prove that the generic pair~$(\Phi_A, \Phi_B) \in L^2(\Sigma_{\tilde{t}}) \oplus L^2(\Sigma_{\tilde{t}})$, that is, a generic initial condition, can be expanded in the set of pairs in~(\ref{initial_cond_set}) with unique coefficients, which proves that such set is a basis of the space of initial conditions. If we write
\begin{equation}
(\Phi_A, \Phi_B) = \sum_n \left[ c_n (\Phi_n, -\rmi \omega_n \Phi_n) + d_n (\Phi_n^*, \rmi \omega_n \Phi_n^*) \right],
\label{expand_ab}
\end{equation}
comparing each component with~(\ref{expand_a_b}) and using~(\ref{expand_p}) we obtain the following set of equations:
\begin{align}
c_m + \sum_n p_{n m} d_n & = a_m, \label{eq_c} \\
\omega_m c_m - \sum_n p_{n m} \omega_n d_n & = \rmi b_m. \label{eq_wc}
\end{align}
Multiplying~(\ref{eq_c}) by~$\omega_m$ and subtracting both equations, we arrive at
\begin{align}
& \sum_n p_{n m} (\omega_n + \omega_m) d_n = \omega_m a_m - \rmi b_m \nonumber \\
\Leftrightarrow & \sum_n p_{n m} d_n = \frac{1}{2} \left( a_m - \rmi \frac{b_m}{\omega_m} \right),
\label{eq_d}
\end{align}
where for obtaining the last equation we have used (\ref{cond_p}). Since~$p_{n m}$ is invertible, the coefficients~$d_n$ exist and are unique. Using~(\ref{eq_c}) we also obtain unique coefficients~$c_n$. This completes the proof.

With respect to the orthonormal character of the bases, using the initial conditions and the normalisation given in~(\ref{normalisation}) one can easily compute the Klein-Gordon inner product~(\ref{scalar_product}) between any two modes of the basis, obtaining:
\begin{align}
\langle \Phi^{[\tilde{t}]}_n, \Phi^{[\tilde{t}]}_m \rangle & = (\omega^{[\tilde{t}]}_n + \omega^{[\tilde{t}]}_m) \int_{\Sigma_{\tilde{t}}} \rmd V_{\tilde{t}}\ \Phi^{[\tilde{t}]}_n(\tilde{t}) \Phi^{[\tilde{t}]}_m(\tilde{t})^* \nonumber \\
& = \delta_{n m},
\label{scalar_product_modes} \\
\langle \Phi^{[\tilde{t}]}_n, (\Phi^{[\tilde{t}]}_m)^* \rangle & = (\omega^{[\tilde{t}]}_n - \omega^{[\tilde{t}]}_m) \int_{\Sigma_{\tilde{t}}} \rmd V_{\tilde{t}}\ \Phi^{[\tilde{t}]}_n(\tilde{t}) \Phi^{[\tilde{t}]}_m(\tilde{t}) \nonumber \\
& = 0,
\label{scalar_product_modes_conj} \\
\langle (\Phi^{[\tilde{t}]}_n)^*, (\Phi^{[\tilde{t}]}_m)^* \rangle & = -(\omega^{[\tilde{t}]}_n + \omega^{[\tilde{t}]}_m) \int_{\Sigma_{\tilde{t}}} \rmd V_{\tilde{t}}\ \Phi^{[\tilde{t}]}_n(\tilde{t})^* \Phi^{[\tilde{t}]}_m(\tilde{t}) \nonumber \\
& = -\delta_{n m};
\label{scalar_product_modes_conj_conj}
\end{align}
where in~(\ref{scalar_product_modes_conj}) we have used that the integral is non-zero only when the eigenvalues coincide. Therefore, we have proven the completeness and orthonormality of the bases of modes that we are constructing for each hypersurface.

\section{Calculations on Bogoliubov transformations}\label{bogo_comp}

\subsection{Detailed calculation of the main result}\label{main_calc}

We consider the Bogoliubov transformation between the modes associated with~$\Sigma_{\tilde{t}_0}$ and with~$\Sigma_{\tilde{t} + \Delta \tilde{t}}$. By composition of Bogoliubov transformations, we have that
\begin{equation}
U(\tilde{t} + \Delta \tilde{t}, \tilde{t}_0) = U(\tilde{t} + \Delta \tilde{t}, \tilde{t}) U(\tilde{t}, \tilde{t}_0).
\label{bogoliubov_composition}
\end{equation}
With the help of this relation, we will find the differential equation in time (\ref{bogoliubov_differential_equation}) for~$U(\tilde{t}, \tilde{t}_0)$. If we differentiate, we get
\begin{align}
\frac{\rmd}{\rmd \tilde{t}} U(\tilde{t}, \tilde{t}_0) & = \left. \frac{\rmd}{\rmd (\Delta \tilde{t})} U(\tilde{t} + \Delta \tilde{t}, \tilde{t}_0) \right|_{\Delta \tilde{t} = 0} \nonumber \\
& = \left. \frac{\rmd}{\rmd (\Delta \tilde{t})} U(\tilde{t} + \Delta \tilde{t}, \tilde{t}) \right|_{\Delta \tilde{t} = 0} U(\tilde{t}, \tilde{t}_0).
\label{bogoliubov_differentiation}
\end{align}
Now we will compute the first factor on the r.h.s. Let us write explicitly the matrix elements of this factor:
\begin{multline}
\left. \frac{\rmd}{\rmd (\Delta \tilde{t})} \alpha_{n m} (\tilde{t} + \Delta \tilde{t}, \tilde{t}) \right|_{\Delta \tilde{t} = 0} = \\
- \rmi \frac{\rmd}{\rmd (\Delta \tilde{t})} \int_{\Sigma_{\tilde{t}}} \rmd V_{\tilde{t}}\ \left[ \Phi^{[\tilde{t} + \Delta \tilde{t}]}_n(\tilde{t}) \left. \partial_t \Phi^{[\tilde{t}]}_m(t)^* \right|_{t=\tilde{t}} \right. \\
\left. \left. - \Phi^{[\tilde{t}]}_m(\tilde{t})^* \left. \partial_t \Phi^{[\tilde{t} + \Delta \tilde{t}]}_n(t) \right|_{t=\tilde{t}} \right] \right|_{\Delta \tilde{t} = 0},
\label{alpha_explicit}
\end{multline}
\begin{multline}
\left. \frac{\rmd}{\rmd (\Delta \tilde{t})} \beta_{n m} (\tilde{t} + \Delta \tilde{t}, \tilde{t}) \right|_{\Delta \tilde{t} = 0} = \\
\rmi \frac{\rmd}{\rmd (\Delta \tilde{t})} \int_{\Sigma_{\tilde{t}}} \rmd V_{\tilde{t}}\ \left[ \Phi^{[\tilde{t} + \Delta \tilde{t}]}_n(\tilde{t}) \left. \partial_t \Phi^{[\tilde{t}]}_m(t) \right|_{t=\tilde{t}} \right. \\
\left. \left. - \Phi^{[\tilde{t}]}_m(\tilde{t}) \left. \partial_t \Phi^{[\tilde{t} + \Delta \tilde{t}]}_n(t) \right|_{t=\tilde{t}} \right] \right|_{\Delta \tilde{t} = 0}.
\label{beta_explicit}
\end{multline}

We need to compute the derivative with respect to~$\Delta \tilde{t}$ of the quantities inside the integrals. For that, we use the \emph{local} evolution in time around~$\tilde{t} + \Delta \tilde{t}$ (and to first order in~$\Delta \tilde{t}$) of~$\Phi^{[\tilde{t} + \Delta \tilde{t}]}_n(t)$, which is given by~(\ref{first_derivative}); and of~$\partial_t \Phi^{[\tilde{t} + \Delta \tilde{t}]}_n(t)$, which we will obtain through the Klein-Gordon equation. Notice that conditions~(\ref{zeroth_derivative}) and~(\ref{first_derivative}), when replaced in the Klein-Gordon equation as written in~(\ref{klein-gordon_2}) and evaluated at~$t=\tilde{t}$, imply that
\begin{align}
\left. \partial^2_t \Phi^{[\tilde{t}]}_n(t) \right|_{t=\tilde{t}} = &\ \rmi \omega^{[\tilde{t}]}_n \left[ \rmi \omega^{[\tilde{t}]}_n + q(\tilde{t}) \right] \Phi^{[\tilde{t}]}_n(\tilde{t}) \nonumber \\
& - [\xi \bar{R} (\tilde{t}) - F(\tilde{t})] \Phi^{[\tilde{t}]}_n(\tilde{t}).
\label{second_derivative}
\end{align}
That is, out of the initial conditions, the Klein-Gordon equation provides the value of the second time derivative of~$\Phi^{[\tilde{t}]}_n(t)$ at~$t=\tilde{t}$ [and \emph{only} at~$t=\tilde{t}$, equation~(\ref{second_derivative}) is evidently \emph{not} a differential equation in time].

Because of relations~(\ref{first_derivative}) and~(\ref{second_derivative}) (replacing $\tilde{t} \to \tilde{t} + \Delta \tilde{t}$), we have that
\begin{multline}
\Phi^{[\tilde{t} + \Delta \tilde{t}]}_n(\tilde{t}) = \Phi^{[\tilde{t} + \Delta \tilde{t}]}_n(\tilde{t} + \Delta \tilde{t}) - \Delta \tilde{t} \left. \partial_t \Phi^{[\tilde{t} + \Delta \tilde{t}]}_n(t) \right|_{t=\tilde{t} + \Delta \tilde{t}} \\
+ O(\Delta \tilde{t})^2 = \Phi^{[\tilde{t} + \Delta \tilde{t}]}_n(\tilde{t} + \Delta \tilde{t}) \left( 1 + \rmi \omega^{[\tilde{t}]}_n \Delta \tilde{t} \right) + O(\Delta \tilde{t})^2, \label{expansion_delta}
\end{multline}
\begin{multline}
\left. \partial_t \Phi^{[\tilde{t} + \Delta \tilde{t}]}_n(t) \right|_{t=\tilde{t}} = \left. \partial_t \Phi^{[\tilde{t} + \Delta \tilde{t}]}_n(t) \right|_{t=\tilde{t} + \Delta \tilde{t}} \\
- \Delta \tilde{t} \left. \partial^2_t \Phi^{[\tilde{t} + \Delta \tilde{t}]}_n(t) \right|_{t=\tilde{t} + \Delta \tilde{t}} + O(\Delta \tilde{t})^2 \\
= -\rmi \omega^{[\tilde{t} + \Delta \tilde{t}]}_n \Phi^{[\tilde{t} + \Delta \tilde{t}]}_n(\tilde{t} + \Delta \tilde{t}) \left\{ 1 + \left[ \rmi \omega^{[\tilde{t}]}_n + q(\tilde{t}) \right] \Delta \tilde{t} \right\} \\
+ \Phi^{[\tilde{t}]}_n(\tilde{t}) [\xi \bar{R} (\tilde{t}) - F(\tilde{t})] \Delta \tilde{t} + O(\Delta \tilde{t})^2. \label{expansion_derivative_delta}
\end{multline}
This is the local evolution of the needed quantities to first order in~$\Delta \tilde{t}$, and therefore we are ready to compute the derivatives in~(\ref{alpha_explicit}) and~(\ref{beta_explicit}). If we plug~(\ref{expansion_delta}) and~(\ref{expansion_derivative_delta}) into~(\ref{alpha_explicit}) and~(\ref{beta_explicit}), we get, with some manipulation,
\begin{align}
\left. \frac{\rmd}{\rmd (\Delta \tilde{t})} \alpha_{n m} (\tilde{t} + \Delta \tilde{t}, \tilde{t}) \right|_{\Delta \tilde{t} = 0} & = \rmi \omega^{[\tilde{t}]}_n \delta_{n m} + \hat{\alpha}_{n m} (\tilde{t}), \label{alpha_split} \\
\left. \frac{\rmd}{\rmd (\Delta \tilde{t})} \beta_{n m} (\tilde{t} + \Delta \tilde{t}, \tilde{t}) \right|_{\Delta \tilde{t} = 0} & = \hat{\beta}_{n m} (\tilde{t});
\label{beta_split}
\end{align}
where the quantities are those defined in~(\ref{alpha_hat}) and~(\ref{beta_hat}). This completes the proof of equation (\ref{bogoliubov_differential_equation}).

\subsection{Bogoliubov identity checking}\label{id}

Any Bogoliubov transformation~$B$ [written in matrix notation as in~(\ref{bogoliubov_matrix})] must satisfy the following identity:
\begin{equation}
B^{-1} = J B^\dagger J, \quad J :=
\left(
\begin{array}{cc}
	I & \\
	 & -I
\end{array}
\right).
\label{bogos_identity}
\end{equation}
The Bogoliubov transformations that we have found in~(\ref{bogoliubov_solution}) and~(\ref{Q_solution}) both have the form~$B = \rme^D$ (the temporal ordering only affects the limits of the integrals and clearly does not play any role). In such a case, noticing that~$J$ is invertible, the condition~(\ref{bogos_identity}) implies the following condition for~$D$:
\begin{equation}
\rme^{-D} = J \rme^{D^\dagger} J = \rme^{J D^\dagger J} \Leftrightarrow D = - J D^\dagger J.
\label{Y_identity}
\end{equation}
Thus, noticing that~$\Omega(\tilde{t})$, $\Theta(\tilde{t})$ and~$A(\tilde{t})$ are diagonal, it is clear that both~$U(\tilde{t}, \tilde{t}_0)$ and~$Q(\tilde{t}, \tilde{t}_0)$ will satisfy the Bogoliubov identity~(\ref{bogos_identity}) if and only if identity~(\ref{Y_identity}) is satisfied for~$D=K(\tilde{t})$. This condition is equivalent to the relations
\begin{equation}
\hat{\alpha}_{n m} (\tilde{t}) = - \hat{\alpha}_{m n} (\tilde{t})^*, \quad \hat{\beta}_{n m} (\tilde{t}) = \hat{\beta}_{m n} (\tilde{t}).
\label{bogos_condition}
\end{equation}

Let us prove these relations by taking the derivative with respect to~$\tilde{t}$ of the inner products given in~(\ref{scalar_product_modes}) and~(\ref{scalar_product_modes_conj}). Since these inner products are constant, with some manipulation we have that
\begin{align}
0 = &\ \frac{\rmd}{\rmd \tilde{t}} \langle \Phi^{[\tilde{t}]}_n, \Phi^{[\tilde{t}]}_m \rangle = \frac{\delta_{n m}}{\omega^{[\tilde{t}]}_n} \frac{\rmd \omega^{[\tilde{t}]}_n}{\rmd \tilde{t}} \nonumber \\
& + (\omega^{[\tilde{t}]}_m + \omega^{[\tilde{t}]}_n) \int_{\Sigma_{\tilde{t}}} \rmd V_{\tilde{t}}\ \bigg\{ \left[ \frac{\rmd}{\rmd \tilde{t}} \Phi^{[\tilde{t}]}_n(\tilde{t}) \right] \Phi^{[\tilde{t}]}_m(\tilde{t})^* \nonumber \\
& + \Phi^{[\tilde{t}]}_n(\tilde{t}) \left[ \frac{\rmd}{\rmd \tilde{t}} \Phi^{[\tilde{t}]}_m(\tilde{t})^* \right] + \Phi^{[\tilde{t}]}_n(\tilde{t}) q(\tilde{t}) \Phi^{[\tilde{t}]}_m(\tilde{t})^* \bigg\},
\label{alpha_derivative} \\
0 = &\ \frac{\rmd}{\rmd \tilde{t}} \langle \Phi^{[\tilde{t}]}_n, (\Phi^{[\tilde{t}]}_m)^* \rangle = \nonumber \\
& \left( \frac{\rmd \omega^{[\tilde{t}]}_m}{\rmd \tilde{t}} - \frac{\rmd \omega^{[\tilde{t}]}_n}{\rmd \tilde{t}} \right) \int_{\Sigma_{\tilde{t}}} \rmd V_{\tilde{t}}\ \Phi^{[\tilde{t}]}_n(\tilde{t}) \Phi^{[\tilde{t}]}_m(\tilde{t}) \nonumber \\
& + (\omega^{[\tilde{t}]}_m - \omega^{[\tilde{t}]}_n) \int_{\Sigma_{\tilde{t}}} \rmd V_{\tilde{t}}\ \bigg\{ \left[ \frac{\rmd}{\rmd \tilde{t}} \Phi^{[\tilde{t}]}_n(\tilde{t}) \right] \Phi^{[\tilde{t}]}_m(\tilde{t}) \nonumber \\
& + \Phi^{[\tilde{t}]}_n(\tilde{t}) \left[ \frac{\rmd}{\rmd \tilde{t}} \Phi^{[\tilde{t}]}_m(\tilde{t}) \right] + \Phi^{[\tilde{t}]}_n(\tilde{t}) q(\tilde{t}) \Phi^{[\tilde{t}]}_m(\tilde{t}) \bigg\}.
\label{beta_derivative}
\end{align}
The last term in each expression appears due to the derivative of~$\rmd V_{\tilde{t}} = \sqrt{h(\tilde{t})} \prod_i \rmd x^i$ [and using~(\ref{change_factor})]. Comparing the identities~(\ref{alpha_derivative}) and~(\ref{beta_derivative}) with the expressions in~(\ref{alpha_hat}) and~(\ref{beta_hat}) it is straightforward to check that the relations in~(\ref{bogos_condition}) hold, and thus the transformations satisfy the Bogoliubov identity~(\ref{bogos_identity}).

\section{Robustness of the results on resonances}\label{invar}

\subsection{Stability under small deviations in the frequency}\label{invar1}

Let us check that resonances still occur when the perturbation frequency slightly deviates from the exact resonant frequency. Consider that a perturbation contains a frequency~$\omega_{\mathrm{p}}$, so that for some pair of modes we get~$\Delta \hat{\alpha}_{n m} (\tilde{t}) = C \rme^{\rmi \omega_{\mathrm{p}} \tilde{t}}$, with~$C$ constant. Consider now that this frequency is very close to the resonant frequency~$\omega^0_n - \omega^0_m$ for the $\alpha$-coefficient between this pair of modes, so that $\omega_{\mathrm{p}} = \omega^0_n - \omega^0_m + \Delta \omega$, with $|\Delta \omega| \lll |\omega^0_n - \omega^0_m|$. From~(\ref{perturbation_alpha}) we have that
\begin{multline}
\alpha_{n m} (\tilde{t}_{\mathrm{f}}, \tilde{t}_0) \approx \varepsilon C \int_{\tilde{t}_0}^{\tilde{t}_{\mathrm{f}}} \rmd \tilde{t}\ \rme^{\rmi \Delta \omega \tilde{t}} \\
= \varepsilon C \frac{2}{\Delta \omega} \sin \left[ \frac{\Delta \omega}{2} (\tilde{t}_{\mathrm{f}} - \tilde{t}_0) \right] \rme^{\rmi \varphi} \approx \varepsilon C (\tilde{t}_{\mathrm{f}} - \tilde{t}_0) \rme^{\rmi \varphi},
\label{freq_perturb}
\end{multline}
where~$\varphi := \Delta \omega (\tilde{t}_{\mathrm{f}} + \tilde{t}_0)/2$ is an irrelevant phase, and we have also assumed that $\tilde{t}_{\mathrm{f}} - \tilde{t}_0 \ll 1/|\Delta \omega|$. Since we also need that $\tilde{t}_{\mathrm{f}} - \tilde{t}_0 \gg 1/|\omega^0_n - \omega^0_m|$ for the resonance to take place, this means that, as far as the frequency is sufficiently close to the resonant one, there is a wide window of time between~$1/|\omega^0_n - \omega^0_m|$ and~$1 / |\Delta \omega|$ in which resonance occurs. Analogous arguments apply to the resonance for the $\beta$-coefficients.

\subsection{Invariance under small changes in the basis of modes}\label{invar2}

In the perturbative regime, the modes with physical interpretation, those that can be used for quantisation, are the resonance modes of the static configuration; that is, the static solutions~$\Phi^0_n$. The only physically meaningful effect that can be described are the resonances between them, since anything else remains as tiny as the perturbation, and therefore as tiny as the degree of arbitrariness in the choice of the solutions to order~$\varepsilon$ that we use to keep track of the evolution. Here, we prove that the results on resonances do not depend on the concrete perturbed bases used to integrate the evolution of the field (as far as they are bases of solutions correctly computed to order~$\varepsilon$).

Consider that, instead of the bases~$\{\Phi^{[\tilde{t}]}_n (t),\Phi^{[\tilde{t}]}_n (t)^*\}$, we used some other bases of solutions~$\{\prescript{1}{}\chi^{[\tilde{t}]}_n (t),\prescript{2}{}\chi^{[\tilde{t}]}_n (t)\}$ which differ just by order~$\varepsilon$ of the original ones (notice that we do not even require that ``half'' of the modes are the complex conjugate of the others). Since both are complete bases of solutions, they have to be related by
\begin{equation}
\left( \begin{array}{c}
	\prescript{1}{}\chi^{[\tilde{t}]}_1 (t) \\
	\ldots \\
	\prescript{2}{}\chi^{[\tilde{t}]}_1 (t) \\
	\ldots
\end{array} \right)
= [I + \varepsilon T (\tilde{t})]
\left( \begin{array}{c}
	\Phi^{[\tilde{t}]}_1 (t) \\
	\ldots \\
	\Phi^{[\tilde{t}]}_1 (t)^* \\
	\ldots
\end{array} \right),
\label{transform_basis}
\end{equation}
where~$T (\tilde{t})$ can be any time-dependent matrix that remains~$O(1)$ in~$\varepsilon$. If we call~$V(\tilde{t},\tilde{t}_0)$ the transformation between the new bases at different times, it is clear that it will be related to the~$U(\tilde{t},\tilde{t}_0)$, which transforms the original bases, by
\begin{equation}
V(\tilde{t},\tilde{t}_0) = [I + \varepsilon T (\tilde{t})] U(\tilde{t},\tilde{t}_0) [I - \varepsilon T (\tilde{t}_0)].
\label{WU_relation}
\end{equation}

Consider now the differential equation~(\ref{bogoliubov_differential_equation}). Taking into account that~$K(\tilde{t})$ has no zeroth order terms in~$\varepsilon$, the corresponding differential equation for~$V(\tilde{t},\tilde{t}_0)$ correct to order~$\varepsilon$ will be
\begin{multline}
\frac{\rmd}{\rmd \tilde{t}} V(\tilde{t}, \tilde{t}_0) = \bigg\{\rmi \Omega(\tilde{t}) + K(\tilde{t}) \\
+ \varepsilon \left[ \frac{\rmd}{\rmd \tilde{t}} T (\tilde{t}) + \rmi \left( T (\tilde{t}) \Omega^0 - \Omega^0 T (\tilde{t}) \right) \right] \bigg\} V(\tilde{t}, \tilde{t}_0),
\label{W_differential_equation}
\end{multline}
with $\Omega^0 := \Omega(\tilde{t}) |_{\varepsilon=0} = \diag (\omega^0_1, \omega^0_2, \ldots, -\omega^0_1, -\omega^0_2, \ldots)$. One can see that, in principle, there is a non-zero contribution to the differential equation due to the change of bases. Let us now write down the matrix~$T (\tilde{t})$ in the block form
\begin{equation}
T (\tilde{t}) :=
\left(
\begin{array}{cc}
	X (\tilde{t}) & Y (\tilde{t}) \\
	Z (\tilde{t}) & W (\tilde{t})
\end{array}
\right).
\label{matrix_T}
\end{equation}
This is just for notational purposes, there is no need that the coefficients satisfy any relation. We consider for example the coefficient~$X_{n m} (\tilde{t})$, which would contribute to the quantity~$\Delta \hat{\alpha}_{n m} (\tilde{t})$ in the new transformation with the amount
\begin{equation}
\frac{\rmd}{\rmd \tilde{t}} X_{n m} (\tilde{t}) - \rmi (\omega^0_n - \omega^0_m) X_{n m} (\tilde{t}),
\label{addition_of_T}
\end{equation}
as one can read out of the new contribution appearing in (\ref{W_differential_equation}). Consider now that the coefficient~$X_{n m} (\tilde{t})$ contains some frequency~$\omega_T$, so that~$X_{n m} (\tilde{t}) = C \rme^{\rmi \omega_T \tilde{t}}$, with~$C$ constant. Then the contribution reads
\begin{equation}
\rmi C (\omega_T - \omega^0_n + \omega^0_m) \rme^{\rmi \omega_T \tilde{t}}.
\label{addition_of_C}
\end{equation}
Finally, when computing the corresponding Bogoliubov coefficient between two different times~$\alpha_{n m} (\tilde{t}, \tilde{t}_0)$ using~(\ref{perturbation_alpha}), the extra contribution due to the change of bases will be
\begin{multline}
\rmi \varepsilon C (\omega_T - \omega^0_n + \omega^0_m) \int_{\tilde{t}_0}^{\tilde{t}_{\mathrm{f}}} \rmd \tilde{t}\ \rme^{\rmi (\omega_T - \omega^0_n + \omega^0_m) \tilde{t}} \\
= \varepsilon C \left[ \rme^{\rmi (\omega_T - \omega^0_n + \omega^0_m) \tilde{t}_{\mathrm{f}}} - \rme^{\rmi (\omega_T - \omega^0_n + \omega^0_m) \tilde{t}_0}\right],
\label{extra_epsilon}
\end{multline}
which remains~$O(\varepsilon)$ at any time, and therefore does not contribute to the resonances. If we had considered the Bogoliubov coefficient as given by the Fourier transform in~(\ref{fourier_alpha}), the conclusion is even more obvious, since the Fourier transform of~(\ref{addition_of_T}) identically vanishes when evaluated at~$\omega^0_n - \omega^0_m$. Analogous arguments apply to the other entries of the transformation.

Notice that, if the transformation~$I + \varepsilon T (\tilde{t})$ is not a Bogoliubov transformation to order~$\varepsilon$, then the Bogoliubov coefficients obtained will not fulfil the Bogoliubov identities to order~$\varepsilon$. This happens because the bases of perturbed solutions used in this case are not orthonormal to order~$\varepsilon$. However, this is completely irrelevant, since these bases are just an intermediate mathematical tool to correctly compute the evolution. The modes in the bases need to be correct solutions to order~$\varepsilon$ and coincide with the static solutions to zeroth order, nothing else. The final Bogoliubov transformation should not be interpreted physically to order~$\varepsilon$, but only beyond it; that is, only for the resonances, which as we have proven remain the same, no matter which transformation (to order~$\varepsilon$) is done to the bases.

It is straightforward to connect the computation done in this Appendix to the claim done in Sect.~\ref{res}, which states that the quantities of the form in~(\ref{no_res_terms_a}) appearing in~$\Delta \hat{\alpha}_{n m} (\tilde{t})$ [respectively of the form in~(\ref{no_res_terms_b}) in $\Delta \hat{\beta}_{n m} (\tilde{t})$] will not contribute to the resonances: As one can see from~(\ref{addition_of_T}), adding or subtracting these quantities is equivalent to an irrelevant change in the choice of bases to order~$\varepsilon$.

\subsection{Further simplification of the expressions}\label{simplif}

Let us obtain expressions~(\ref{super_simpl_a}-\ref{superoperator}) from~(\ref{fourier_alpha}) and (\ref{fourier_beta}). First, we write the operator~$\hat{\mathscr{O}}(\tilde{t})$ to first order in~$\varepsilon$ as
\begin{equation}
\hat{\mathscr{O}}(\tilde{t}) \approx \hat{\mathscr{O}}^0 + \varepsilon \Delta \hat{\mathscr{O}} (\tilde{t}).
\label{operator_eps}
\end{equation}
Zeroth and first orders in~$\varepsilon$ of equation~(\ref{zeroth_derivative}) then read, respectively,
\begin{align}
\hat{\mathscr{O}}^0 \Phi^0_n & = (\omega^0_n)^2 \Phi^0_n, \label{zero_zero} \\
\hat{\mathscr{O}}^0 \Delta \Phi^{[\tilde{t}]}_n(\tilde{t}) + \Delta \hat{\mathscr{O}} (\tilde{t}) \Phi^0_n & = (\omega^0_n)^2 \Delta \Phi^{[\tilde{t}]}_n(\tilde{t}) + 2 \omega^0_n \Delta \omega^{[\tilde{t}]}_n \Phi^0_n. \label{zero_one}
\end{align}
Using these two expressions and the fact that $\hat{\mathscr{O}}^0$ is self-adjoint, we can write
\begin{align}
(\omega^0_m)^2 & \int_{\Sigma^0} \rmd V^0\ \Delta \Phi^{[\tilde{t}]}_n(\tilde{t}) (\Phi^0_m)^* \nonumber \\
= & \int_{\Sigma^0} \rmd V^0\ \Delta \Phi^{[\tilde{t}]}_n(\tilde{t}) [\hat{\mathscr{O}}^0 (\Phi^0_m)^*] \nonumber \\
= & \int_{\Sigma^0} \rmd V^0\ [\hat{\mathscr{O}}^0 \Delta \Phi^{[\tilde{t}]}_n(\tilde{t})] (\Phi^0_m)^* \nonumber \\
= & - \int_{\Sigma^0} \rmd V^0\ [\Delta \hat{\mathscr{O}} (\tilde{t}) \Phi^0_n] (\Phi^0_m)^* \nonumber \\
& + (\omega^0_n)^2 \int_{\Sigma^0} \rmd V^0\ \Delta \Phi^{[\tilde{t}]}_n(\tilde{t}) (\Phi^0_m)^* + \Delta \omega^{[\tilde{t}]}_n \delta_{n m}; \label{la_prueba}
\end{align}
which implies
\begin{multline}
\rmi [(\omega^0_n)^2 - (\omega^0_m)^2] \int_{\Sigma^0} \rmd V^0\ \Delta \Phi^{[\tilde{t}]}_n(\tilde{t}) (\Phi^0_m)^* + \rmi \Delta \omega^{[\tilde{t}]}_n \delta_{n m} = \\
\rmi \int_{\Sigma^0} \rmd V^0\ [\Delta \hat{\mathscr{O}} (\tilde{t}) \Phi^0_n] (\Phi^0_m)^*.
\end{multline}
Finally, we notice that, because of the equivalence relation with respect to resonances, any diagonal term in~$\Delta \hat{\alpha}_{n m} (\tilde{t})$ is always meaningless. We can thus ignore the second term in the l.h.s., obtaining~(\ref{super_simpl_a}) and~(\ref{superoperator}) for~$\Delta \hat{\alpha}_{n m} (\tilde{t})$. An identical derivation with the replacement $(\Phi^0_m)^* \to \Phi^0_m$ yields~(\ref{super_simpl_b}) for~$\Delta \hat{\beta}_{n m} (\tilde{t})$, in which case the diagonal terms are relevant.

\section{Summary of formulae for the application of the method}\label{recipes}

In this Appendix, we provide Tables~\ref{summary} and~\ref{summary2} with all the formulae necessary in order to apply the method to a concrete problem. As we indicated in the examples in Subsect.\ \ref{gw} and Sect.\ \ref{cosmo}, once the expressions for the method have been found, there is no need to consider the explicit evolution in time of the modes constructed any more. Therefore, the notation for the ``time label'' of the different modes can be simplified replacing~$\tilde{t} \to t$. In the tables we use this simplification.

\begin{table*}
\caption{Summary of formulae for the application of the method (part~1)}
\label{summary}
\begin{tabular*}{\textwidth}{@{}lc}
\hline
\textbf{Computation of modes and eigenvalues}\\
\hline
Eigenvalue equation (\ref{zeroth_derivative})&
$\begin{aligned}
\hat{\mathscr{O}}(t) \Phi^{[t]}_n (t) = (\omega^{[t]}_n)^2 \Phi^{[t]}_n (t);
\end{aligned}$
\vspace{0.2cm}
\\
with the operator (\ref{operator})&
$\begin{aligned}
\hat{\mathscr{O}}(t) = &\ - \nabla_{h(t)}^2 + \xi R^h(t) + m^2 + F(t), \\
\text{with}\ F(t)\ & \text{such that}\ \hat{\mathscr{O}}(t)\ \text{is positive definite};
\end{aligned}$
\vspace{0.2cm}
\\
(possibly) boundary conditions (\ref{dirichlet_static}-\ref{robin_static})&
$\begin{aligned}
\Phi^{[t]}_n(t, \vec{x}) & = 0, \quad \vec{x} \in \partial \Sigma_{t} & \text{(Dirichlet)}, \\
\vec{n} \cdot \nabla_{h(t)} \Phi^{[t]}_n(t, \vec{x}) & = 0, \quad \vec{x} \in \partial \Sigma_{t} & \text{(Neumann)}, \\
\vec{n} \cdot \nabla_{h(t)} \Phi^{[t]}_n(t, \vec{x}) + \gamma \Phi^{[t]}_n(t, \vec{x}) & = 0, \quad \vec{x} \in \partial \Sigma_{t} & \text{(Robin)};
\end{aligned}$
\vspace{0.2cm}
\\
and orthonormalisation condition (\ref{normalisation})&
$\begin{aligned}
\int_{\Sigma_{t}} \rmd V_{t}\ \Phi^{[t]}_n(t)\Phi^{[t]}_m(t)^* = \frac{\delta_{n m}}{2 \omega^{[t]}_n}.
\end{aligned}$\\
\hline
\textbf{Non-perturbative regime}\\
\hline
Quantities~$\hat{\alpha}$ and~$\hat{\beta}$ (\ref{alpha_hat}, \ref{beta_hat})&
$\begin{aligned}
\hat{\alpha}_{n m} (t) = &\ (\omega^{[t]}_m + \omega^{[t]}_n) \int_{\Sigma_{t}} \rmd V_{t}\ \left[ \frac{\rmd}{\rmd t} \Phi^{[t]}_n(t) \right] \Phi^{[t]}_m(t)^*\\
& + \int_{\Sigma_{t}} \rmd V_{t}\ \Phi^{[t]}_n(t) \left[\omega^{[t]}_n q(t) + \rmi \xi \bar{R} (t) \right] \Phi^{[t]}_m(t)^*\\
& + \frac{\delta_{n m}}{2 \omega^{[t]}_n} \left[ \frac{\rmd \omega^{[t]}_n}{\rmd t} - \rmi F(t) \right],\\
\hat{\beta}_{n m} (t) = &\ (\omega^{[t]}_m - \omega^{[t]}_n) \int_{\Sigma_{t}} \rmd V_{t}\ \left[ \frac{\rmd}{\rmd t} \Phi^{[t]}_n(t) \right] \Phi^{[t]}_m(t)\\
& - \int_{\Sigma_{t}} \rmd V_{t}\ \Phi^{[t]}_n(t) \left[\omega^{[t]}_n q(t) + \rmi \xi \bar{R} (t) \right] \Phi^{[t]}_m(t)\\
& - \left[ \frac{\rmd \omega^{[t]}_n}{\rmd t} - \rmi F(t) \right] \int_{\Sigma_{t}} \rmd V_{t}\ \Phi^{[t]}_n(t) \Phi^{[t]}_m(t);
\end{aligned}$
\vspace{0.2cm}
\\
with (\ref{change_factor}, \ref{bar_scalar})&
$\begin{aligned}
q(t) & = \partial_t \log \sqrt{h(t)},\\
\bar{R}(t) & = 2 \partial_t q(t) + q(t)^2 - [\partial_t h^{i j} (t)] [\partial_t h_{i j} (t)]/4.
\end{aligned}$\\
\hline
Bogoliubov coefficients (with the trivial phase)\\
\hline
Differential equation~(\ref{bogoliubov_differential_equation})&
$\begin{aligned}
\frac{\rmd}{\rmd t} U(t, t_0) = \left[\rmi \Omega(t) + K(t) \right] U(t, t_0);
\end{aligned}$
\vspace{0.2cm}
\\
with (\ref{transform_matrices})&
$\begin{aligned}
\Omega (t) & = \diag (\omega^{[t]}_1, \omega^{[t]}_2, \ldots, -\omega^{[t]}_1, -\omega^{[t]}_2, \ldots),\\
K(t) & =
\left(
\begin{array}{cc}
	\hat{\alpha} (t) & \hat{\beta} (t) \\
	\hat{\beta} (t)^* & \hat{\alpha} (t)^*
\end{array}
\right).
\end{aligned}$
\vspace{0.2cm}
\\
Formal solution (\ref{bogoliubov_solution})&
$\begin{aligned}
U(t_{\mathrm{f}}, t_0) = \mathscr{T} \exp \left\{ \int_{t_0}^{t_{\mathrm{f}}} \rmd t \left[ \rmi \Omega(t) + K(t) \right] \right\}.
\end{aligned}$\\
\hline
Bogoliubov coefficients (without the trivial phase)\\
\hline
Differential equation~(\ref{Q_differential_equation})&
$\begin{aligned}
\frac{\rmd}{\rmd t} Q(t, t_0) = \Theta(t)^* \bar{K}(t) \Theta(t) Q(t, t_0);
\end{aligned}$
\vspace{0.2cm}
\\
with (\ref{exponential_omega})&
$\begin{aligned}
\Theta(t) & = \exp \left\{ \int^{t} \rmd t' [\rmi \Omega (t') + A(t')] \right\},\\
\bar{K}(t) & = K(t) - A(t),\\
A(t) & = \diag(\hat{\alpha}_{1 1}, \hat{\alpha}_{2 2}, \ldots, -\hat{\alpha}_{1 1}, -\hat{\alpha}_{2 2}, \ldots).
\end{aligned}$
\vspace{0.2cm}
\\
Formal solution (\ref{Q_solution})&
$\begin{aligned}
Q(t_{\mathrm{f}}, t_0) = \mathscr{T} \exp \left[ \int_{t_0}^{t_{\mathrm{f}}} \rmd t\ \Theta(t)^* \bar{K}(t) \Theta(t) \right].
\end{aligned}$\\
\hline
\end{tabular*}
\end{table*}

\newpage

\begin{table*}
\caption{Summary of formulae for the application of the method (part~2)}
\label{summary2}
\begin{tabular*}{\textwidth}{@{}lc}
\hline
\textbf{Perturbative regime}\\
\hline
Quantities~$\Delta \hat{\alpha}$ and~$\Delta \hat{\beta}$\\
\hline
Using the modes computed to first order in~$\varepsilon$ (\ref{alpha_hat_simpl}, \ref{beta_hat_simpl})&
$\begin{aligned}
\Delta \hat{\alpha}_{n m} (t) \equiv &\ \rmi [(\omega^0_n)^2 - (\omega^0_m)^2] \int_{\Sigma^0} \rmd V^0\ \Delta \Phi^{[t]}_n(t) (\Phi^0_m)^*\\
& + \int_{\Sigma^0} \rmd V^0\ \Phi^0_n \left[\omega^0_n \Delta q(t) + \rmi \xi \Delta \bar{R}(t) \right] (\Phi^0_m)^*,\\
\Delta \hat{\beta}_{n m} (t) \equiv &\ -\rmi [(\omega^0_n)^2 - (\omega^0_m)^2] \int_{\Sigma^0} \rmd V^0\ \Delta \Phi^{[t]}_n(t) \Phi^0_m\\
& - \int_{\Sigma^0} \rmd V^0\ \Phi^0_n \left[\omega^0_n \Delta q(t) + \rmi \xi \Delta \bar{R}(t) \right] \Phi^0_m\\
& - [2 \rmi \omega^0_n \Delta \omega^{[t]}_n - \rmi \Delta F(t) ] \int_{\Sigma^0} \rmd V^0\ \Phi^0_n \Phi^0_m.
\end{aligned}$
\vspace{0.2cm}
\\
Using the modes solution to the static problem (\ref{super_simpl_a}, \ref{super_simpl_b})&
$\begin{aligned}
\Delta \hat{\alpha}_{n m} (t) & \equiv \int_{\Sigma^0} \rmd V^0\ [\hat{\Delta}(t) \Phi^0_n] (\Phi^0_m)^*,\\
\Delta \hat{\beta}_{n m} (t) & \equiv - \int_{\Sigma^0} \rmd V^0\ [\hat{\Delta}(t) \Phi^0_n] \Phi^0_m;
\end{aligned}$
\vspace{0.2cm}
\\
with (\ref{superoperator}, \ref{operator_eps})&
$\begin{aligned}
\hat{\Delta}(t) \Phi^0_n & = \left[ \rmi \Delta \hat{\mathscr{O}} (t) + \omega^0_n \Delta q(t) + \rmi \xi \Delta \bar{R}(t) - \rmi \Delta F(t) \right] \Phi^0_n,\\
\Delta \hat{\mathscr{O}} (t) & = \partial_\varepsilon \hat{\mathscr{O}}(t) |_{\varepsilon=0}.
\end{aligned}$\\
\hline
Bogoliubov coefficients\\
\hline
Explicit time evolution (resonances) (\ref{perturbation_alpha}, \ref{perturbation_beta})&
$\begin{aligned}
\alpha_{n n} (t_{\mathrm{f}}, t_0) & \approx 1;\\
\alpha_{n m} (t_{\mathrm{f}}, t_0) & \approx \varepsilon \int_{t_0}^{t_{\mathrm{f}}} \rmd t\ \rme^{-\rmi (\omega^0_n - \omega^0_m) t} \Delta \hat{\alpha}_{n m} (t), \quad n \neq m;\\
\beta_{n m} (t_{\mathrm{f}}, t_0) & \approx \varepsilon \int_{t_0}^{t_{\mathrm{f}}} \rmd t\ \rme^{-\rmi (\omega^0_n + \omega^0_m) t} \Delta \hat{\beta}_{n m} (t).
\end{aligned}$
\vspace{0.2cm}
\\
Asymptotic values using Fourier transforms (\ref{fourier_alpha}, \ref{fourier_beta})&
$\begin{aligned}
\alpha_{n n} (-\infty,\infty) & \approx 1;\\
\alpha_{n m} (-\infty,\infty) & \approx \varepsilon \sqrt{2 \pi}\ \mathscr{F} [\Delta \hat{\alpha}_{n m}] (\omega^0_n - \omega^0_m), \quad n \neq m;\\
\beta_{n m} (-\infty,\infty) & \approx \varepsilon \sqrt{2 \pi}\ \mathscr{F} [\Delta \hat{\beta}_{n m}] (\omega^0_n + \omega^0_m).
\end{aligned}$\\
\hline
\end{tabular*}
\end{table*}

\bibliographystyle{spphys} 
\bibliography{ContBogos} 

\end{document}